\documentclass{PoS}

\newcommand{\gev}{\,\, \mathrm{GeV}}
\newcommand{\tb}{\tan\beta}
\newcommand{\ssi}{\sigma^{\rm SI}_p}

\def\beq{\begin{equation}}
\def\eeq{\end{equation}}

\title{Supersymmetric Dark Matter after Run I at the LHC: From a TeV to a PeV}

\ShortTitle{From a TeV to a PeV}

\author{\speaker{Keith A. Olive}\\
       William I. Fine Theoretical Physics Institute, School of Physics and Astronomy,\\
University of Minnesota, Minneapolis, MN 55455,\,USA\\
        E-mail: \email{olive@umn.edu}}

\abstract{The absence of low energy supersymmetry in run I data at the LHC
has pushed the nominal scale for supersymmetry beyond a TeV. While this is consistent
with the discovery of the Higgs boson at $\approx 125$ GeV, simple models 
with scalar and gaugino mass universality are being pushed into corners of parameter space.
Some possibilities within the constrained minimal supersymmetric extension of the standard 
model (with four parameters) are discussed along with a one parameter extension in which 
the Higgs soft masses are non-universal. Also discussed are 2-, 3-, and 4-parameter
versions of pure gravity mediated models with a wino, Higgsino, or bino LSP respectively. 
}

\FullConference{18th International Conference From the Planck Scale to the Electroweak Scale \\
		25-29 May 2015\\
		Ioannina, Greece }

\begin{document}

\section{Introduction}
The ATLAS and CMS experiments did not discover supersymmetry
during run I at the LHC \cite{ATLAS20,CMS20} and instead have pushed the supersymmetric
scale towards a TeV or beyond. So why still consider supersymmetry as a low energy extension of the 
Standard Model? Clearly as the energy scale for supersymmetry is pushed up,
its ability to solve the hierarchy problem 'natually' becomes more difficult. However,
the large hierarchy problem associated with Grand Unified Theories (GUTs) or Planck scale is
still nicely resolved by supersymmetry \cite{hier} even if some residual fine tuning is necessary.
In addition, supersymmetry still aids the problems of gauge coupling unification \cite{Ellis:1990zq}; the 
stabilization of the electroweak vacuum \cite{Ellis:2000ig}; radiative electroweak symmetry breaking 
\cite{ewsb}; and dark matter \cite{ehnos}.  In addition simple supersymmetric models
generally predict that the Higgs boson mass should be $< 130$ GeV \cite{Ellis:1990nz} which has been 
confirmed at the LHC \cite{lhch}. 

Here, I will consider several classes of supersymmetric models related to the well-studied
constrained minimal supersymmetric Standard Model (CMSSM) \cite{funnel,cmssm,efgo,cmssmwmap,eo6,ehow+,elos,eelnos}. 
The CMSSM is a 4-parameter theory defined at the GUT scale by a universal
gaugino mass, $m_{1/2}$, a universal scalar mass, $m_0$, a universal trilinear mass term, $A_0$,
and the ratio of the two Higgs vacuum expectation values, $\tan \beta$. In addition, 
one must specify the sign of the Higgs mixing mass, $\mu$. The gravitino mass, $m_{3/2}$,
is usually assumed to be heavy enough to be ignorable in CMSSM phenomenology.
One can define a more restrictive theory related more closely to minimal supergravity (mSUGRA) 
\cite{Fetal,acn,bfs}. In the CMSSM, one uses the conditions derived
by the minimization of the Higgs potential after radiative electroweak symmetry breaking to solve for
$\mu$ and the bilinear mass term $B_0$ (or equivalently $\mu$ and the Higgs pseudo scalar mass, $m_A$)
for fixed $\tan \beta$.
In these very constrained models \cite{bfs,vcmssm}, the relation $A_0 = B_0 + m_0$ is imposed
and $\tan \beta$ is no longer a free parameter. In addition, one should set the scalar mass
equal to the gravitino mass ($m_0 = m_{3/2}$) and this becomes a 3-parameter theory.
One could relax the CMSSM by allowing the two Higgs soft masses to differ from the universal
scalar mass, producing either a 5- or 6-parameter theory, depending on whether 
the two Higgs soft masses are equal (NUHM1) \cite{nuhm1,eosknuhm} or not (NUHM2) \cite{nuhm2,eosknuhm}. Finally, I will discuss a very restrictive 2-parameter theory known as
pure gravity mediation \cite{pgm,pgm2,ArkaniHamed:2012gw,eioy,eioy2,eo} defined by only $m_{3/2}$ and $\tan \beta$ (and the sign of $\mu$), as well as some of its 1- and 2-parameter extensions.

\section{The CMSSM before and after Run I at the LHC}

\begin{figure}
  \includegraphics[height=.35\textwidth]{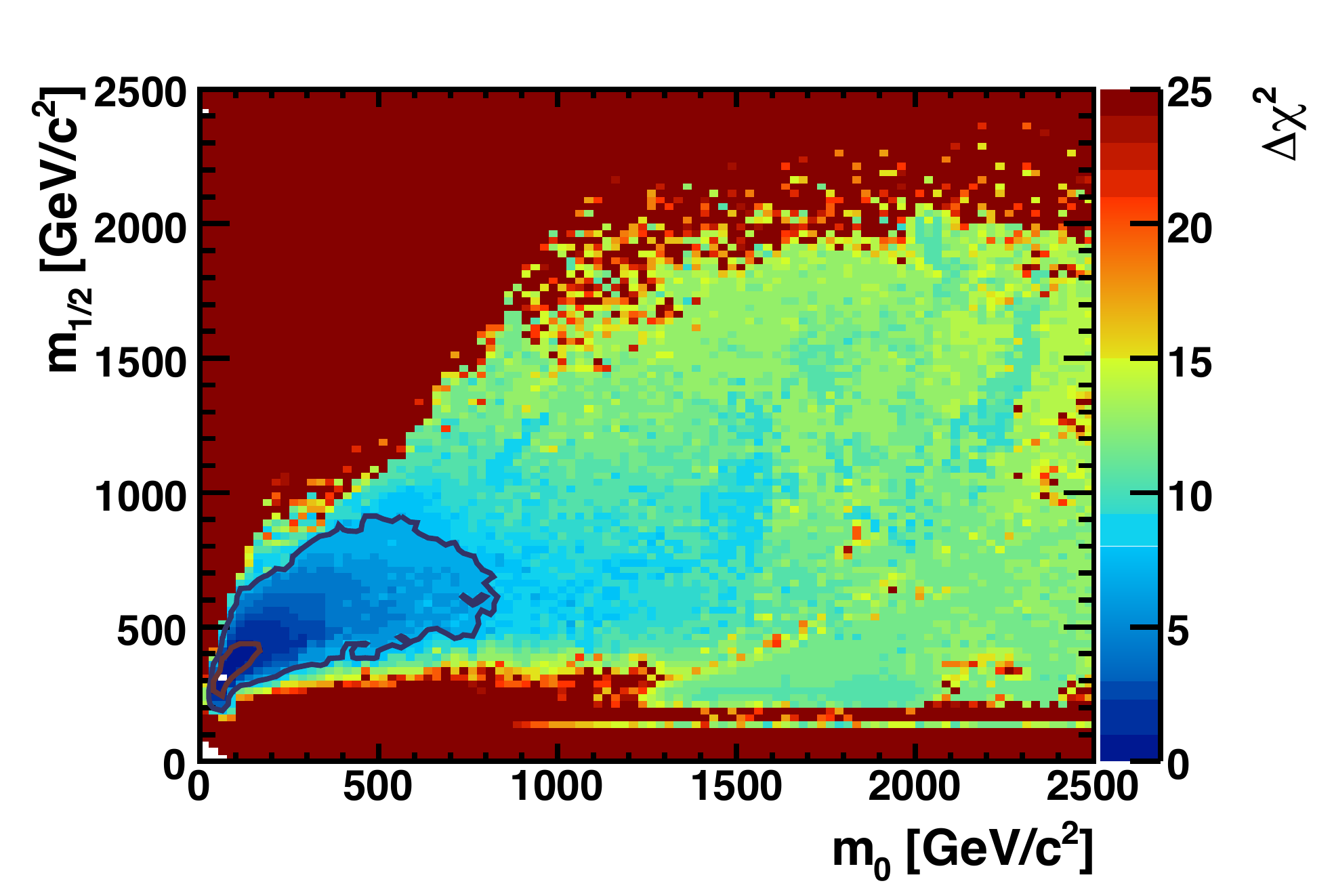}
   \includegraphics[height=.35\textwidth]{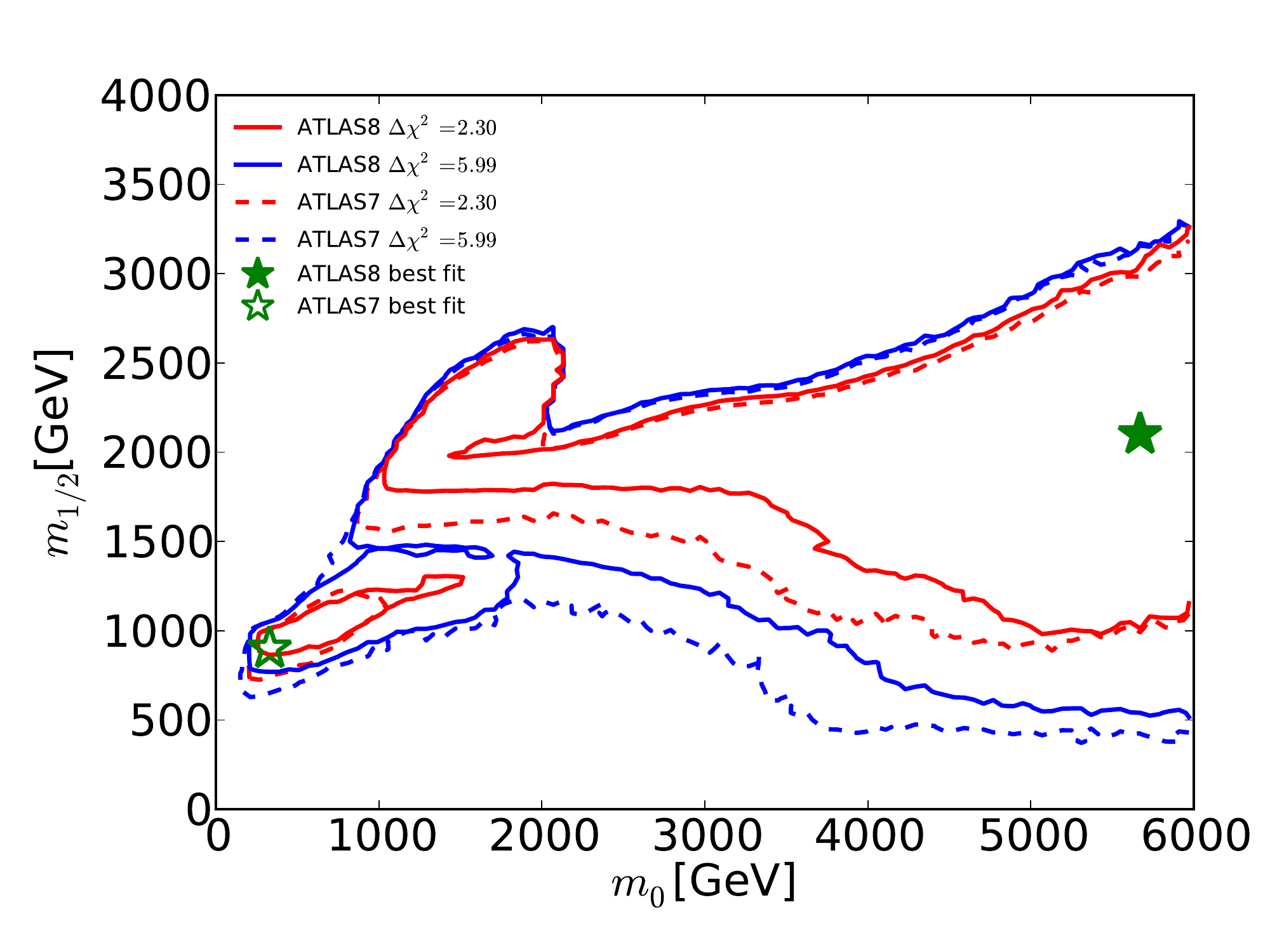}
  \caption{The $\Delta\chi^2$ functions in the $(m_0, m_{1/2})$ planes for
  the CMSSM from a mastercode frequentist analysis. The pre-LHC result is shown in the 
  left panel \cite{mc3}. Red and blue contours correspond to 68\% and 95\% CL contours and the best fit point is 
  depicted by a white dot. The post-LHC result is shown in the right panel \cite{mc9}. Dashed contours
  correspond to the analysis using 7 TeV data at 5 fb$^{-1}$ and solid curves to 8 TeV data at 20 fb$^{-1}$.  }
  \label{mcm0m12}
\end{figure}

Prior to Run I of the LHC, there was considerable excitement about the prospect for discovering 
supersymmetry as supersymmetric models such as the CMSSM
provided definite improvements to low energy precision phenomenology.
This can be seen in the left panel of Fig. \ref{mcm0m12} which shows the 
results of mastercode \cite{mc3,mc} - a 
frequentist Markov Chain Monte Carlo analysis of low energy experimental observables in
the context of supersymmetry. The figure shows the color coded values of $\Delta \chi^2$
relative to the best fit point shown by the white dot at low $m_{1/2}$ and low $m_0$.
Marginalization over $A_0$ and $\tan \beta$ was performed to produce this $(m_0, m_{1/2})$ plane.
The best-fit CMSSM point lies at
$m_0 = 60 \gev$,  $m_{1/2} = 310 \gev$,  $A_0 = 130 \gev$, $\tb = 11$
yielding the overall $\chi^2/{\rm N_{\rm dof}} = 20.6/19$ (36\% probability) 
with $m_h = 114.2$ GeV. Recall that this was a pre-LHC prediction and uses no LHC data.
Rather it is based on a wide array of low energy observables including $(g_\mu -2)$, $M_W$, 
$B \to \tau \nu$, $b \to s \gamma$, the LEP limit on the Higgs mass, forward-backward asymmetries
among others (for a full list of observables used see \cite{mc3}). The relatively low value of $m_h$
was a common prediction of MSSM models \cite{Ellis:1990nz}. Indeed a dedicated scan for the distribution 
of Higgs masses in the CMSSM was made in \cite{ENOS} which found that when 
all phenomenological constraints (with or without $(g_\mu -2)$) are included, all models yielded
$m_h \le 128$ GeV. When $(g_\mu -2)$ is included, only models with $m_h < 126$ GeV were found.
Note that the scan sampled scalar and gaugino masses only out to 2 TeV.

\begin{figure}
  \includegraphics[height=.5\textwidth]{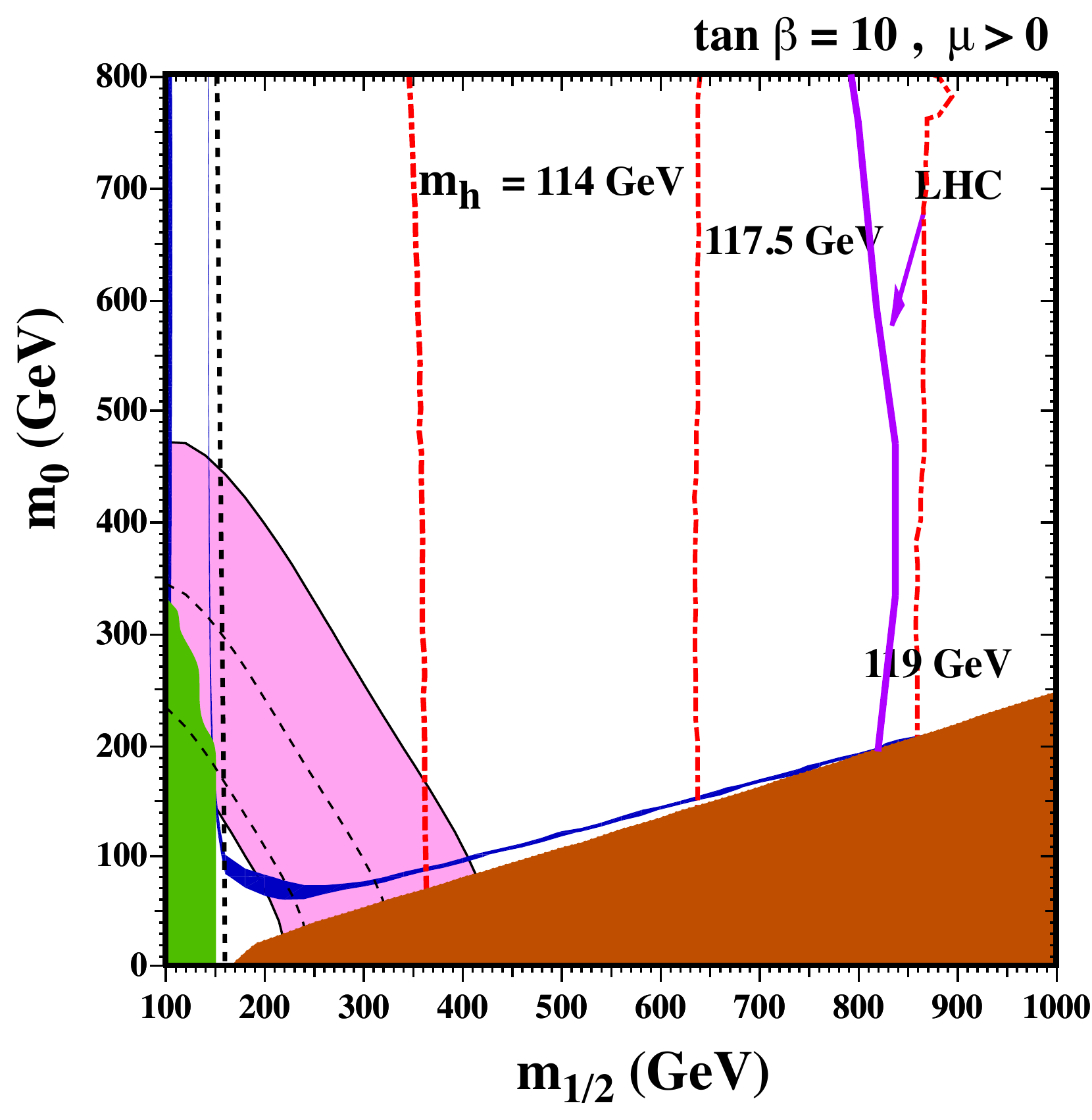}
   \includegraphics[height=.5\textwidth]{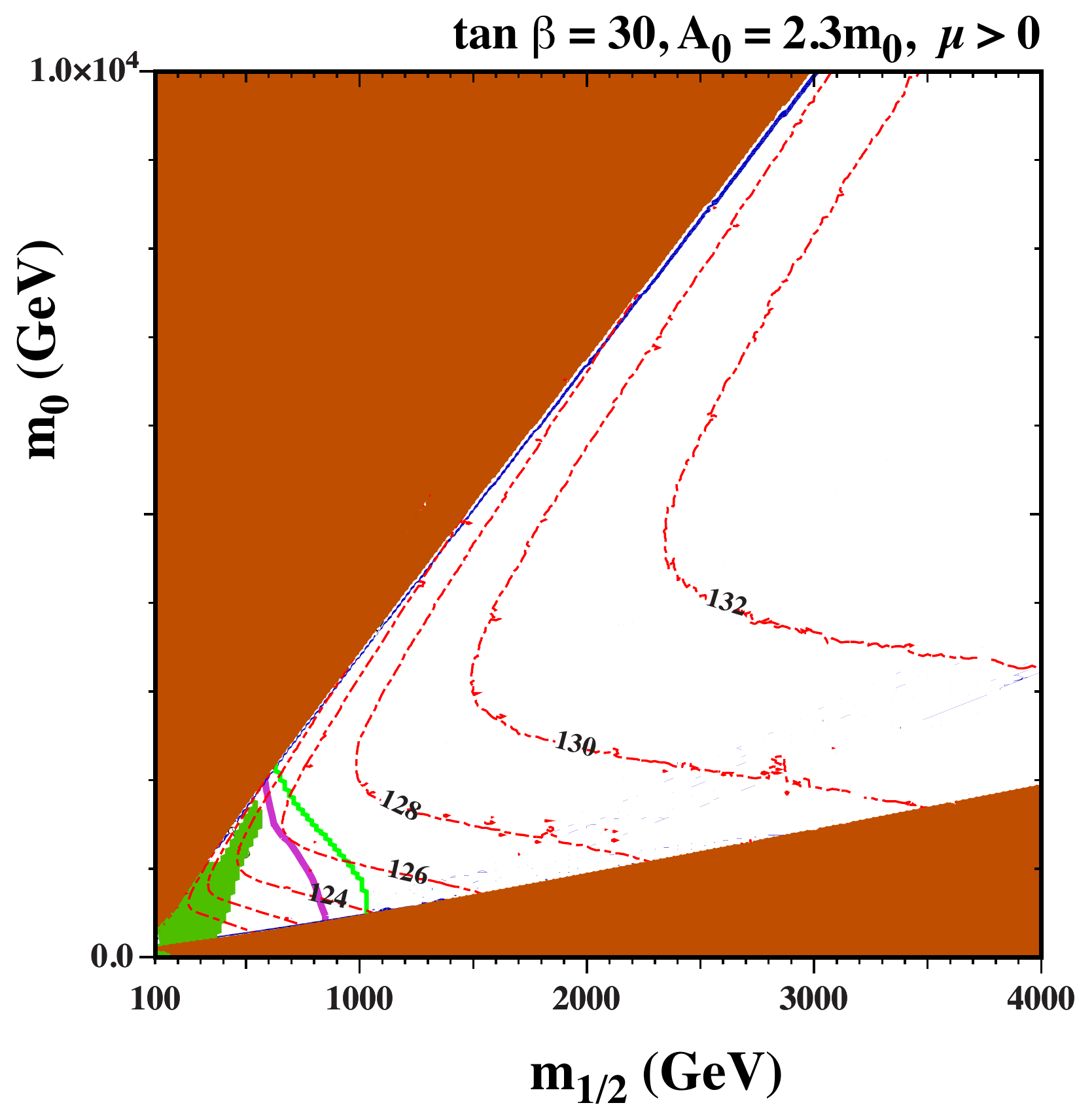}
  \caption{(left) The $(m_{1/2}, m_0)$ plane for $\tan \beta = 10$ and  $\mu > 0$, 
assuming $A_0 = 0$. The near-vertical (red)
dot-dashed lines are the contours of constant $m_h$, and the near-vertical (black) dashed
line is the contour $m_{\chi^\pm} = 104$~GeV.  The medium (dark
green) shaded region is excluded by $b \to s
\gamma$, and the dark (blue) shaded area is the cosmologically
preferred region. In the dark
(brick red) shaded region, the LSP is the charged ${\tilde \tau}_1$. The
region preferred by $(g_\mu -2)$ at the 2-$\sigma$
level, is shaded (pink) and bounded by solid black lines, with dashed
lines indicating the 1-$\sigma$ range.
The curve marked LHC show the 95\% CL exclusion region (to the left of the curve)
for the final run I LHC sparticle searches. (right) The $(m_{1/2}, m_0)$ plane for $\tan \beta = 30$ and  $\mu > 0$, 
assuming $A_0 = 2.3 m_0$.  The red shaded region in the upper left is excluded because there the stop, ${\tilde t}_1$ is the LSP. The green curve to the right of the LHC bound is due to $B \to \mu^+ \mu^-$ and
also excludes everything to its left. }
  \label{m0m121}
\end{figure}

The apparent robustness for the success of the CMSSM can also be seen in the left panel of 
Fig. \ref{m0m121} where an example of a $(m_{1/2}, m_0)$ plane with fixed $A_0 = 0$ and 
$\tb = 10$ is shown. The red shaded region in the lower right corned is excluded as
the lighter stau is the lightest supersymmetric particle (LSP). Above that region,
one sees the stau coannihilation strip where the relic density is in agreement with that 
determined by Planck \cite{Planck15}. The relic density there is held in check through
coannihilations between the bino and the lighter stau \cite{stau}.    
Also shown in the figure are some phenomenological
constraints from the lack of detection of charginos \cite{LEPsusy}, 
as well as constraints from $b \to s \gamma$ \cite{bsgex}
and $(g_\mu - 2)$ \cite{newBNL}.  The purple curve shows the lower limit on $m_{1/2}$ from the LHC \cite{ATLAS20,CMS20}. 
One immediately sees how the promise of the CMSSM evaporated
when the LHC results set in.

\begin{figure}
  \includegraphics[height=.35\textwidth]{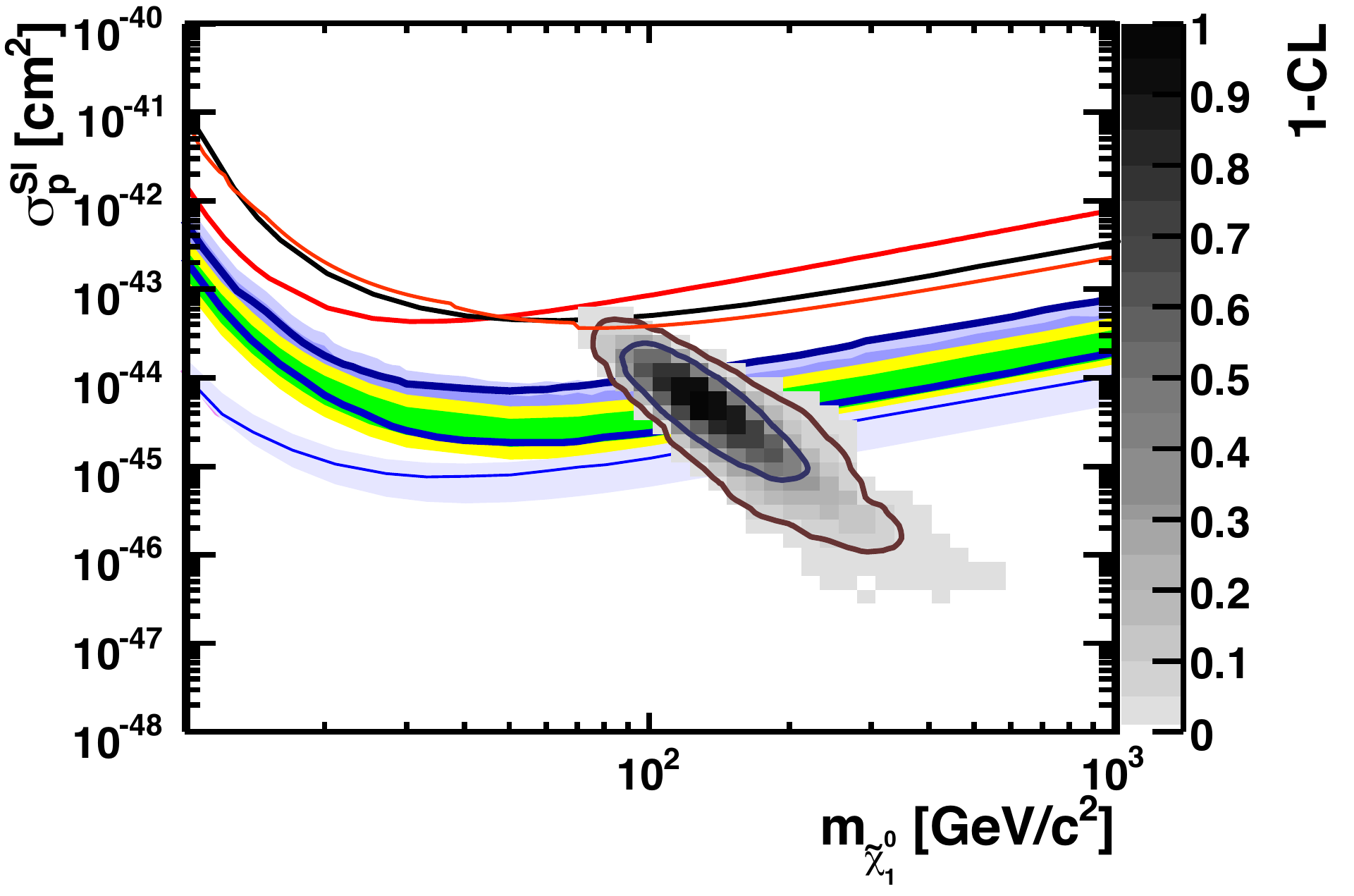}
   \includegraphics[height=.40\textwidth]{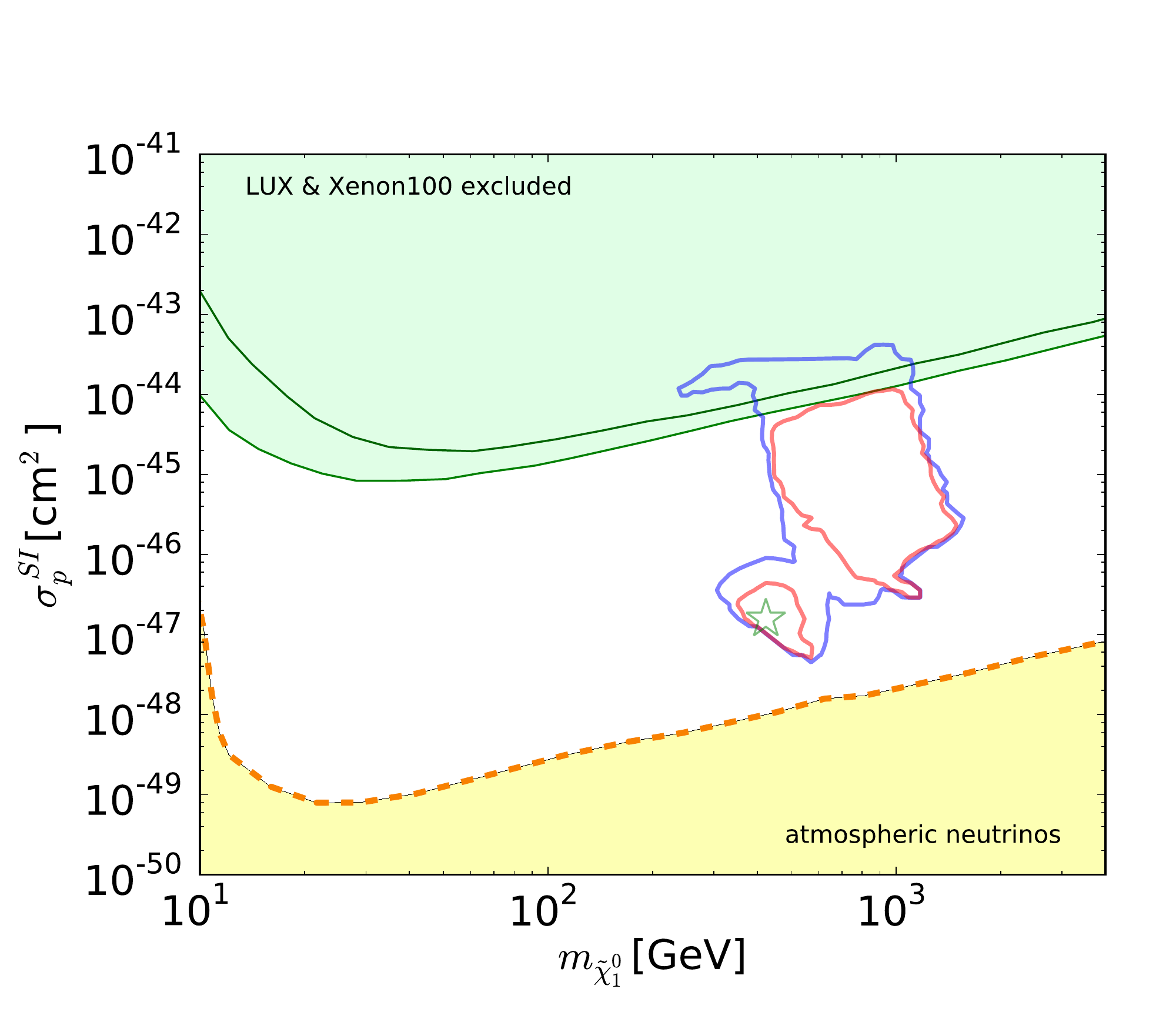}
  \caption{(left) The pre-LHC prediction for the spin-independent DM
scattering cross section, $\ssi$,  versus $m_\chi$
in the CMSSM. The solid lines
are the pre-LHC experimental upper limits from CDMS~\cite{CDMS} and XENON10\cite{Xe10},
while the bands are the more recent limits from XENON100 \cite{xe100100,xe100} and LUX \cite{lux}.   
(right) The post-run I likelihood contours for $\ssi$.  The shaded region at low cross section 
represents the neutrino background for direct detection experiments \cite{nuback}. }
  \label{mcssi}
\end{figure}

This optimism of discovering supersymmetry spread to the prospects of discovering dark matter
in direct detection experiments. The left panel of Fig.~\ref{mcssi} displays the 
pre-LHC preferred range of the
spin-independent DM scattering cross section $\ssi$
(calculated here assuming an optimistic $\pi$-N scattering term
$\Sigma_N = 64$~MeV) as a function of $m_\chi$ \cite{mc3}. 
We see that
the expected range of $\ssi$ lies just below the then present
experimental upper limits (solid lines) \cite{CDMS,Xe10}.
As one can see from the successive lower upper limits from later experiments \cite{xe100100,xe100,lux}
shown by the bands, these pre-LHC values for the elastic scattering cross section showed great 
promise for discovery.

Subsequent to the final run at the LHC, the picture looked very different.
In the right panel of Fig. \ref{mcm0m12}, the post-Run I likelihood contours in the $(m_0, m_{1/2})$ plane \cite{mc9}
are shown using both the 7 TeV results at 5 fb$^{-1}$ (dashed) \cite{ATLAS5} and 8 TeV results at
20 fb$^{-1}$ (solid) \cite{ATLAS20}. At 7 TeV, the previous best fit at $(m_0, m_{1/2})$ = (60,n310) GeV
has now moved to (340, 910) GeV with $A_0 = 2670$ GeV and $\tb = 12$. The $\chi^2/N_{\rm dof}$
is now 32.6/23 (8.8\% probability). This result includes the measurement of the Higgs mass of $m_h = 125.7 \pm 0.4$ (at the time) \cite{MH-ATLAS} and the best fit point is shown by the open star. 
The likelihood function is in fact very shallow. The best fit point based on the 8 TeV data is shown by the
filled star at (5650,2100) GeV with $A_0 = -780$ GeV and $\tb = 51$. The $\chi^2/N_{\rm dof}$
is now  increased again to 35.1/23 (5.1\% probability). However there is a local minimum at lower masses
at (670, 1040) GeV with $A_0 = 3440$ GeV and $\tb = 21$.  In this case, $\chi^2/N_{\rm dof}$
is 35.8/23 (4.3\% probability). These results may be compared with the Standard Model fit which yields
$\chi^2/N_{\rm dof}$
of 36.5/24 (5.0\% probability), which of course ignores the fact that there is no dark matter candidate in the Standard 
Model. 

An example of the $(m_{1/12}, m_0)$ plane with fixed $\tb = 30$ and $A_0/m_0 = 2.3$ for $\mu > 0$
is shown in the right panel of Fig. \ref{m0m121}. In addition to a stau LSP region (as in the left panel),
there is also a stop LSP region indicated by the shaded region in the upper left of the panel.
The area between these two have a neutralino (bino) LSP. 
The dark blue strips running near the boundaries of these regions have a relic LSP density in accord
with that determined by Planck~\cite{Planck15}, though the widths of these dark matter strips have been enhanced for visibility. The strip near the
boundary of the upper left wedge is due to stop coannihilation \cite{stop,eoz,raza}, 
and that near the boundary of the lower
right wedge is due to stau coannihilation \cite{stau}. 
Stop coannihilation strips typically extend to much larger values of $m_{1/2}$ than
stau coannihilation strips (which extend to $\sim 1000$ GeV), indeed to much larger values of $m_{1/2}$
than those displayed in the figure, reaching as far as 8000 GeV in this case \cite{eoz}.

\begin{figure}
 \includegraphics[height=.35\textwidth]{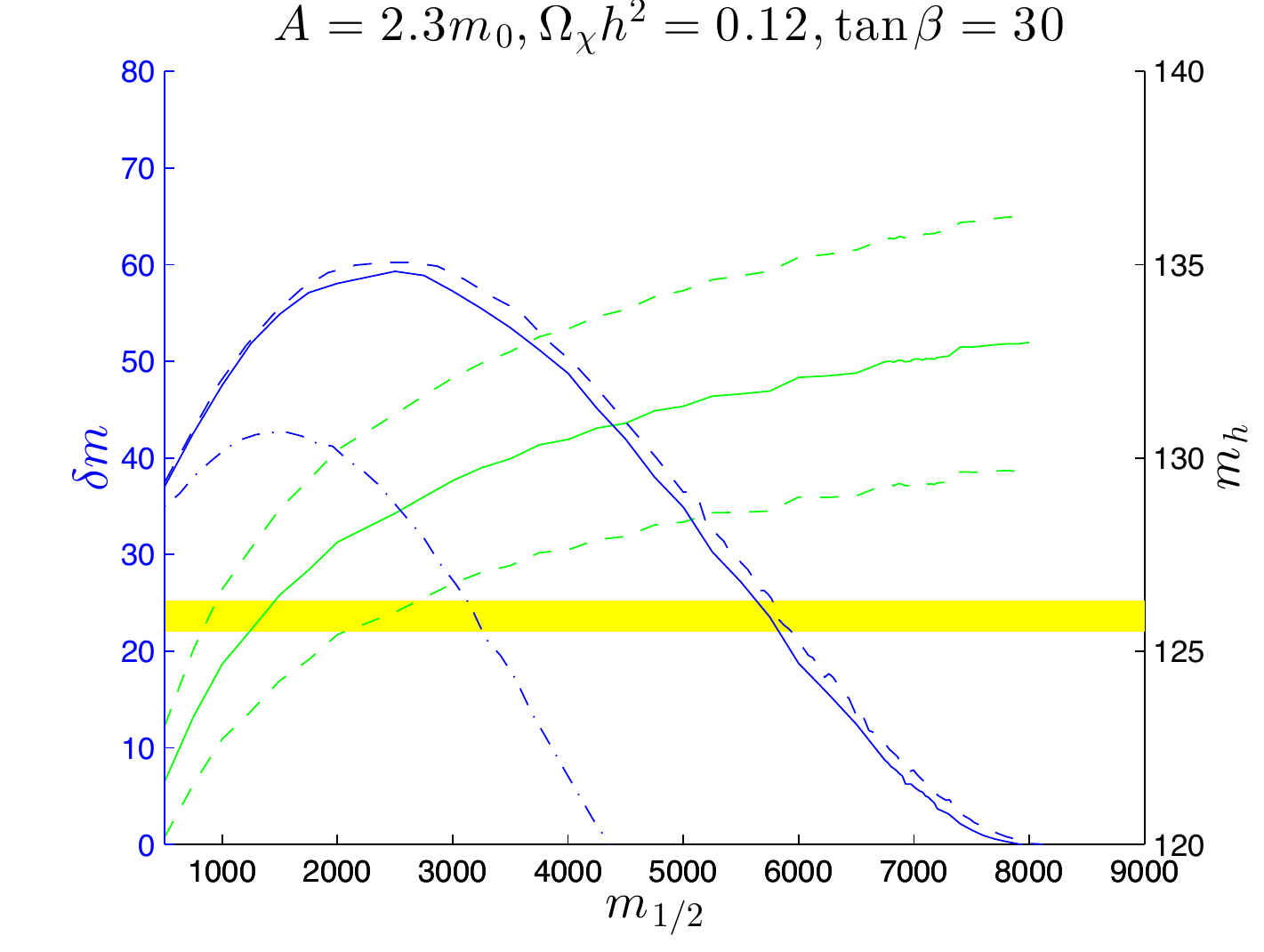}
  \includegraphics[height=.35\textwidth]{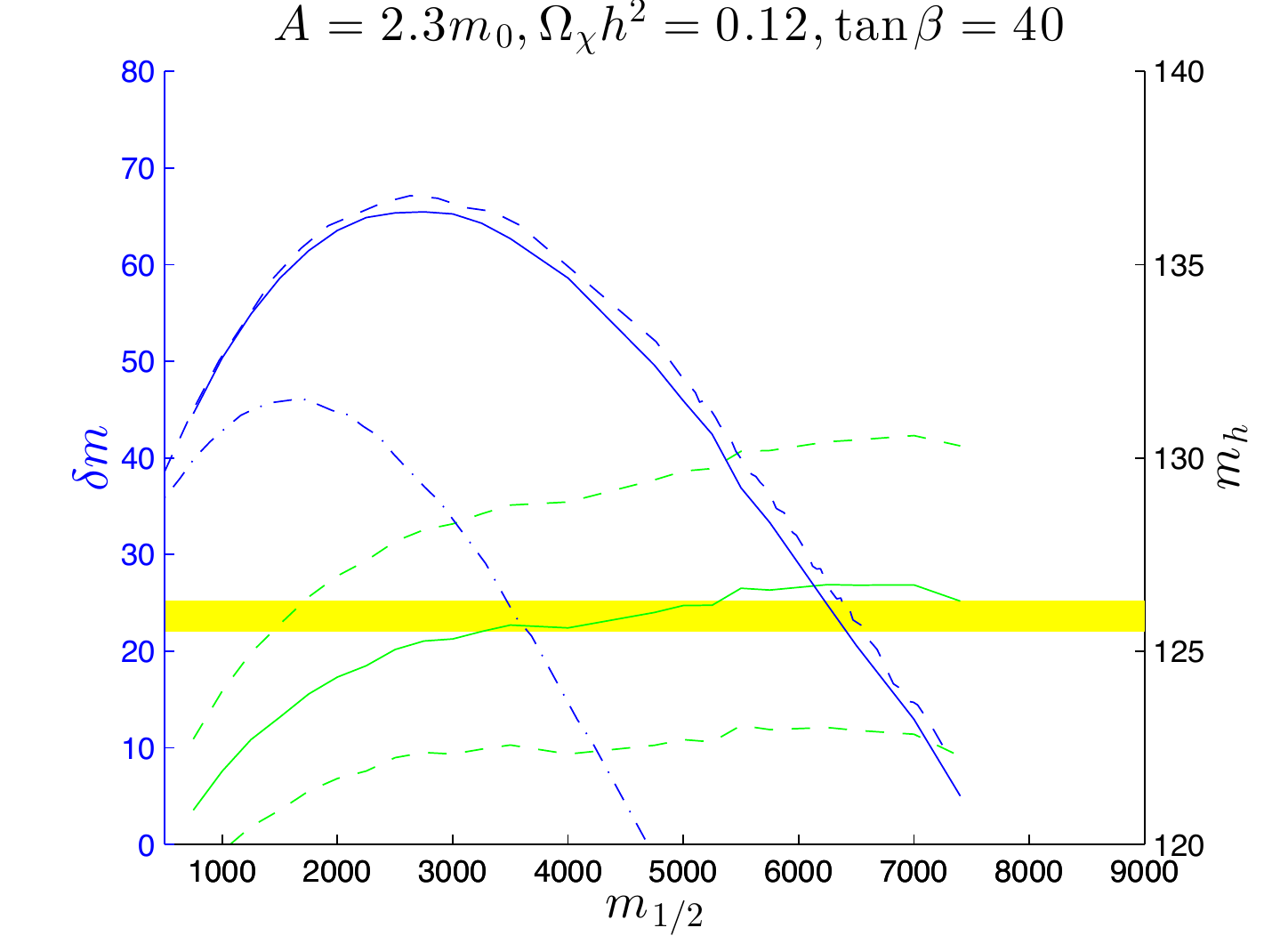}
  \caption{The mass difference $\delta m=m_{\tilde t_1}-m_\chi$ and the Higgs mass $m_h$
(all masses in GeV units) as
functions of $m_{1/2}$ along the stop coannihilation strip where $\Omega_{\chi} h^2=0.12$, 
for $\tan\beta=30$ (left) and $\tb = 40$ (right) both with $A=2.3 \, m_{0}$. 
The solid blue lines show the values of $\delta m$
incorporating the Sommerfeld corrections. The dashed blue lines show $\delta m$ with 
$\Omega_{\chi} h^2=0.125$ and the dot dashed blues line show $\delta m$ without the Sommerfeld
correction. The green lines show the values of $m_h$, with
the dashed lines representing the uncertainty range given by {\tt FeynHiggs~2.10.0}. 
The yellow bands indicate the current experimental value of $m_h$ and its uncertainty. }
  \label{stopstrip}
\end{figure}

The extent of the stop strip for this case is seen in the left panel of Fig. \ref{stopstrip} which shows
the stop-neutralino mass difference as a function of $m_{1/2}$ for fixed relic density \cite{eoz}.
The endpoint is at approximately $(m_{1/2}, m_0) = (8000, 26900)$ 
with a neutralino mass of about 4000 GeV. 
Past the endpoint, even for $m_{\tilde t_1} = m_\chi$, the relic density is too large. 
Also shown in Fig. \ref{stopstrip} is the value of the Higgs mass along the strip (green lines).
The dashed lines correspond to the uncertainty in the Higgs mass. The Higgs mass (and uncertainties) 
here and in previous figures are calculated using {\tt FeynHiggs~2.10.0} \cite{fh}.
The Higgs mass at the endpoint is too large ($m_h = 133$ GeV) in this case, but is acceptable at the 
endpoint when $\tb = 40$ as seen in the right panel of Fig. \ref{stopstrip}. In the latter case, the endpoint is found at $(m_{1/2}, m_0) = (7600, 38600)$ with $m_\chi \approx 3900$ GeV and $m_h = 126.2$ GeV.

Also shown in the right panel of Fig. \ref{m0m121} is the exclusion region from  $b \to s \gamma$
decay \cite{bsgex}, and the green solid line is the 95 \% CL constraint from the measured rate of $B_s \to \mu^+ \mu^-$ decay \cite{bmm}. As in the left panel, the purple curve shows the position of 
the run I constraint \cite{ATLAS20,CMS20}. 

As one can see from the right panel of Fig. \ref{mcssi}, there is still hope for direct detection experiments though the new best fit point implies a cross section of $\sim 10^{-47}$ cm$^2$, nearly two orders of magnitude below the current upper bound.  As commented on previously, the likelihood function
is rather flat between $10^{-47}$ cm$^2 \lesssim \ssi\ \lesssim 10^{-45}$~cm$^2$.
Note that in this case, we have adopted $\Sigma_{\pi N} = 50 \pm 
7$ MeV. In addition to the model results,
 the 90\% CL upper limits on $\ssi\ $ given by the XENON100 and LUX
 experiments~\cite{xe100,lux} are also displayed, as is the level of the atmospheric neutrino background~\cite{nuback}.
 
 \section{The NUHM}

Going beyond the CMSSM offers additional possibilities for low energy supersymmetry and dark matter \cite{elos,eelnos}.
For example, the Higgs soft masses may differ from the squark and slepton masses at the universal
input scale \cite{nonu,efgo,nuhm1,eosknuhm,nuhm2} as in the NUHM. In SU(5) or SO(10) 
GUTs, the Higgs fields are placed in representations distinct from the matter fields and it is a logical
possibility that their soft masses may differ at the GUT scale. In an SO(10) theory, both Higgs doublets
are found in the same representation and we might expect that $m_1 = m_2$ as in the NUHM1.
In contrast, in SU(5), they originate in different representations and we might expect $m_1 \ne m_2$
as in the NUHM2.  In the CMSSM, setting $m_1 = m_2 = m_0$ allows one to use the Higgs minimization
conditions to solve for the two Higgs expectation values. However, in practice, these are usually 
chosen as inputs (using $M_Z$ and $\tan \beta$ as surrogates) and instead the minimization conditions are
used to solve for the $\mu$ parameter and the bilinear mass term, $B$. 
In the NUHM, with $m_1$ and/or $m_2$
as free parameters, one can trade these instead to choose either or both $\mu$ and/or $B$ (or the Higgs 
pseudo scalar mass, $m_A$ which is related to $B \mu$) as free parameters. 
Additional freedom is also obtained in CMSSM-like models when the scale of universality is
taken to be below the GUT scale as in subGUT models \cite{subGUT,elos,eelnos}, or
above the GUT scale as in superGUT models \cite{superGUT,dlmmo}, but these will not be discussed here. 

\begin{figure}
 \includegraphics[height=.5\textwidth]{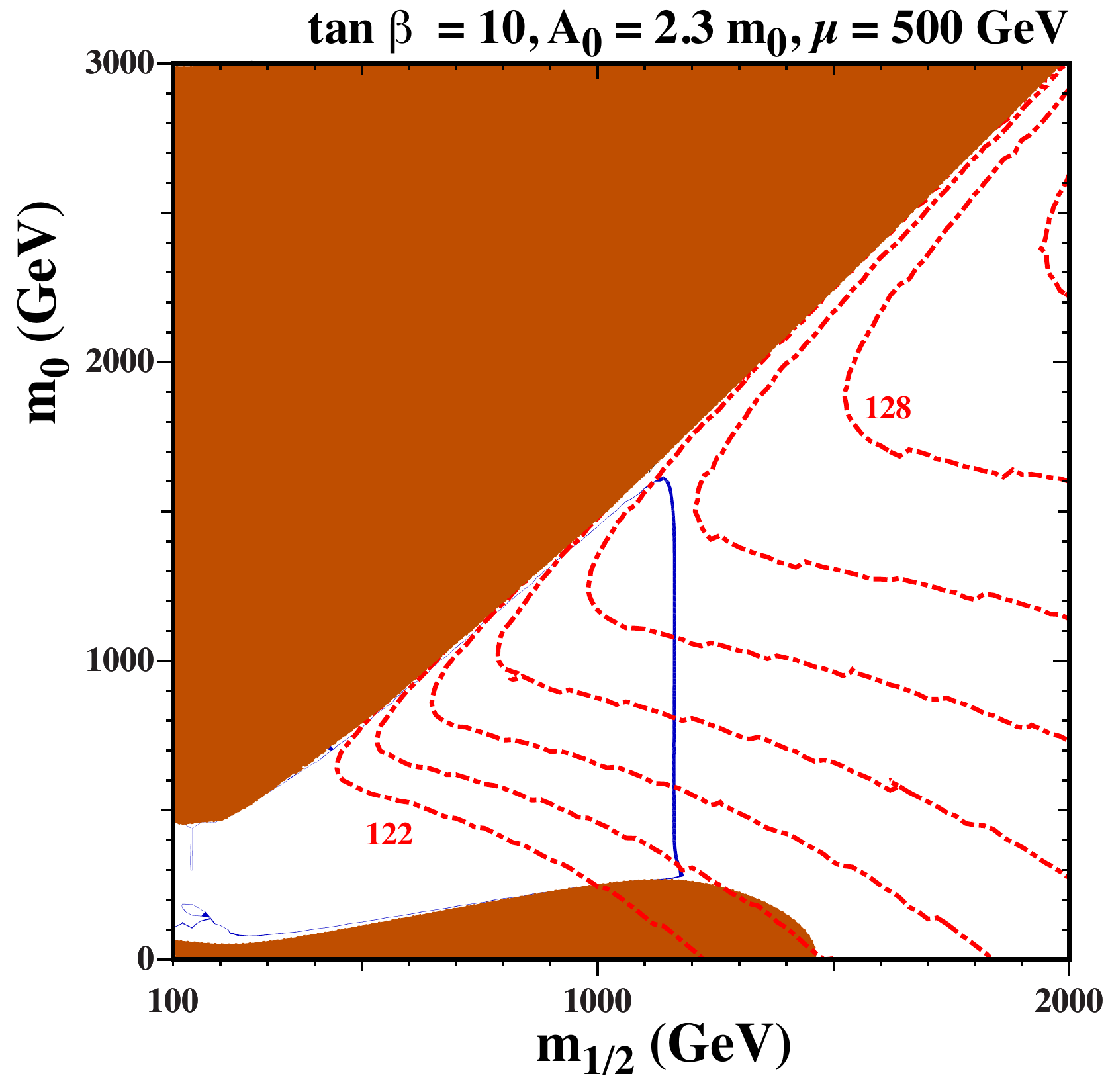}
  \includegraphics[height=.5\textwidth]{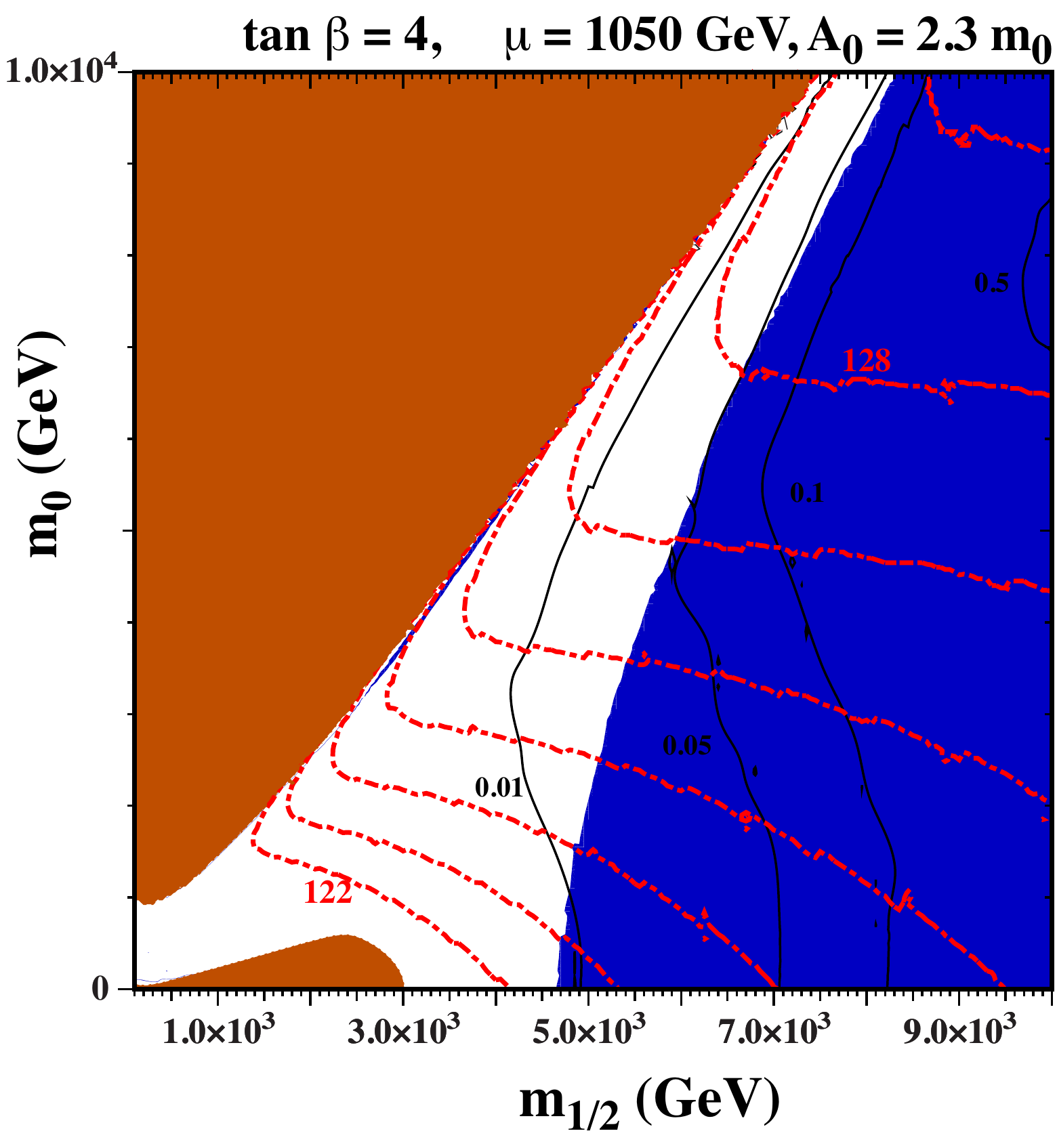}
  \caption{The NUHM1  $(m_{1/2}, m_0)$ planes for $\tan \beta = 10$ and $\mu = 500$ GeV (left) and $\tan \beta = 4$ and $\mu = 1050$ GeV (right). 
In both panels $A_0 = 2.3 m_0$. The shading and contour types are as in Fig.~2. The solid black curves in the right panel correspond to contours of constant proton lifetime in units of $10^{35}$ yrs. }
  \label{nuhm1}
\end{figure}

In Fig. \ref{nuhm1}, two examples of $(m_{1/2}, m_0)$ NUHM1 planes are shown with fixed $\mu = 500$ GeV
with $\tan \beta = 10$ (left panel) and $\mu = 1050$ GeV with $\tan \beta = 4$ (right panel). In both $A_0 = 2.3 m_0$. For the smaller value of $\mu$, there is a thin vertical transition strip at about $m_{1/2} = 1200$ GeV.
To the left of the strip, the LSP is a bino and away from the shaded regions the relic density is too large.
Since $\mu$ is fixed, as one increases the gaugino masses, the LSP becomes more Higgsino-like.
As such, new annihilation channels open up and the relic density drops. To the right of the 
strip, the LSP is mostly Higgsino and because its mass is close to $\mu$, the relic density is 
too small for this choice of $\mu$. As in the CMSSM, there is a shaded region
where the LSP is a stop in the upper left of the panel and a stau in the lower left. 
There are barely visible stop/stau coannihilation
strips that runs close to the stop/stau LSP boundary. Here, there is no enhancement of the relic density
strip and along the strip, the relic density is within 3$\sigma$ of the Planck result. 
The Higgs mass along the transition strip runs from
about 123 to 127 GeV in agreement with the experimental value when taking into account the 
theoretical uncertainty in the calculated value. 

In the right panel of of Fig. \ref{nuhm1}, we see a large region where relic density is in agreement with Planck. As before, the LSP is bino-like to the left of the transition region and the relic density is too high expect for the
extremely thin strips along the stop and stau LSP areas. In this case as the gaugino mass is increased,
the LSP once again becomes more Higgsino-like, however asymptotically, at large $m_{1/2}$, the Higgsino 
mass tends toward $\simeq 1.1$ TeV where Higgsinos provide the correct dark matter density (this is of course
tied to the choice of $\mu = 1050$ GeV) \cite{osi}. Indeed, this region would extend infinitely far to the right. 
As one can see, there is a significant area where the Higgsino provides the correct relic density
with an acceptable Higgs mass. Also shown in this panel are contours of constant 
proton lifetime (in units of $10^{35}$ yrs) assuming a minimal SU(5) GUT \cite{eelnos}. The experimental
limit $\tau (p \to K^+ \bar{\nu}) > 5.9\times 10^{33}$~years
\cite{Abe:2014mwa} would exclude points to the left of the curve labelled 0.05 or require
a non-minimal GUT for which the calculated lifetime satisfies the bound.

\section{Pure Gravity Mediation}

In the remainder of this contribution, I will concentrate on an even more restrictive set of models
known as pure gravity mediation (PGM) \cite{pgm,pgm2,ArkaniHamed:2012gw,eioy,eioy2,eo} 
which as noted in the introduction can be defined by as little as two free parameters. 
In the minimal model of PGM,  one assumes a flat K\"ahler potential, and there are no tree level sources for
either gaugino masses or $A$-terms.
At one-loop, gaugino masses and $A$-terms are generated through anomalies \cite{anom} and
one expects $m_{1/2}, A_0 \ll m_0$ in these models, reminiscent of split supersymmetry
\cite{split}. So far, $m_0 = m_{3/2}$ (as in mSUGRA models \cite{bfs}) is the only free parameter.  
Radiative electroweak symmetry breaking (EWSB) can be incorporated into the
model at the expense of one additional parameter, $c_H$, associated with a Giuduce-Masiero-like term \cite{gm,ikyy,dmmo}
 which amounts to including
\beq
\Delta K = c_H H_1 H_2  + h.c. \, ,
\label{gmk}
\eeq
in the K\"ahler potential. Here, $c_H$ is a constant and the expressions for $\mu$ and $B$ are modified at the input scale taken here to be $M_{GUT}$,
\begin{eqnarray}
 \mu &=& \mu_0 + c_H m_{3/2}\ ,
 \label{eq:mu0}
 \\
  B\mu &=&  \mu_0 (A_0 - m_{3/2}) + 2 c_H m_{3/2}^2\ ,
   \label{eq:Bmu0}
\end{eqnarray}
where $\mu_0$ is the $\mu$-term of the superpotential.
 One can also easily trade $c_H$ for $\tan \beta$, leaving the theory to be defined by $m_{3/2}$, $\tan \beta$ (with $c_H$ determined by the Higgs minimization conditions) and the sign of the $\mu$
term. A similar particle spectrum was also derived in models with strong moduli stabilization
\cite{dmmo,lmo,dlmmo}.

To realize such a model, consider the breaking of supersymmetry using a Polonyi superpotential \cite{pol}
\beq
W= \mu^2(Z+\nu),
\label{W}
\eeq
where the parameter $\nu$ is adjusted so that the cosmological constant vanishes at the supersymmetry breaking minimum.
The K\"ahler potential includes a stabilization term \cite{dine},
\beq\label{K}
K = Z\bar{Z}-\frac{(Z\bar{Z})^2}{\Lambda^2},
\eeq
where it is assumed that the mass scale $\Lambda\ll 1$.  
If we write $Z$ in terms of
its real and imaginary parts, $Z=\frac{1}{\sqrt{2}}(z+i\chi)$
 the supersymmetry breaking Minkowski minimum is found to be real and located at
\beq\label{z_min}
\langle z\rangle_{\rm Min}\simeq \frac{\Lambda^2}{\sqrt{6}}\ , \ \ \langle \chi\rangle =0\ , \ \ \nu\simeq \frac{1}{\sqrt{3}}\ .
\eeq
for $\Lambda\ll 1$. The supersymmetry breaking mass scale given by the gravitino mass is
\beq
m_{3/2}=\langle e^{K/2}W\rangle \simeq\mu^2/\sqrt{3} \, ,
\eeq
whereas the mass squared of both $z$ and $\chi$ are
\beq
m_{z,\chi}^2\simeq \frac{12\,m_{3/2}^2}{\Lambda^2}\gg m_{3/2}^2 .
\label{mz}
\eeq
Thus, for $\Lambda \ll 1$, we obtain a hierarchy between the modulus, $Z$,  and the gravitino. 
As a result we obtain the following:
\begin{eqnarray}
m_0^2 & = & m_{3/2}^2 \\
A_0 & = & {1 \over 2} m_{3/2} \Lambda^2 + {\rm anomalies} \\
B_0 & = & A_0 - m_0 \approx -m_0 \\
M_i & = & {\rm anomalies}
\end{eqnarray}
The dominant source for gaugino masses comes from the one-loop anomaly mediated contributions\cite{anom}, which are given by
\begin{eqnarray}
    M_{1} &=&
    \frac{33}{5} \frac{g_{1}^{2}}{16 \pi^{2}}
    m_{3/2}\ ,
    \label{eq:M1} \\
    M_{2} &=&
    \frac{g_{2}^{2}}{16 \pi^{2}} m_{3/2}  \ ,
        \label{eq:M2}     \\
    M_{3} &=&  -3 \frac{g_3^2}{16\pi^2} m_{3/2}\ .
    \label{eq:M3}
\end{eqnarray}
Here, the subscripts of $M_a$, $(a=1,2,3)$, correspond to the Standard Model gauge groups. 
As a result, in PGM, the gaugino masses are much smaller than the scalar masses and are non-universal
even in the minimal model. Furthermore, because of the loop suppression to these masses,
we are forced to consider rather large scalar masses. 

In summary, the minimal model of PGM is much simpler than mSUGRA and has
only two fundamental parameters, $m_0$ and $c_H$ (or $\tan \beta$) once 
the vacuum conditions of  electroweak symmetry breaking are used. 
$m_{3/2}$ must lie in the range of  $\mathcal{O}$(100) TeV so that
the predicted chargino mass satisfies the LEP bound, the Higgs mass
agrees with the LHC measurement, and gaugino masses
are within the reach of the LHC experiments.
With only these two parameters, 
the model leads to a very successful phenomenology;
\begin{itemize}
\item The sfermion and  gravitino have masses $\mathcal{O}$(100) TeV.
\item The higgsino and the heavier Higgs boson also have masses $\mathcal{O}$(100) TeV.
\item The gaugino masses are in the range of  hundreds to thousands of GeV.
\item The LSP is the neutral wino which is nearly degenerate with the charged wino.
\item The lightest Higgs boson mass is consistent with the observed Higgs-like boson,
i.e. $m_h \simeq 125$--$126$\,GeV.
\end{itemize}

\begin{figure}[h]
 \includegraphics[height=.5\textwidth]{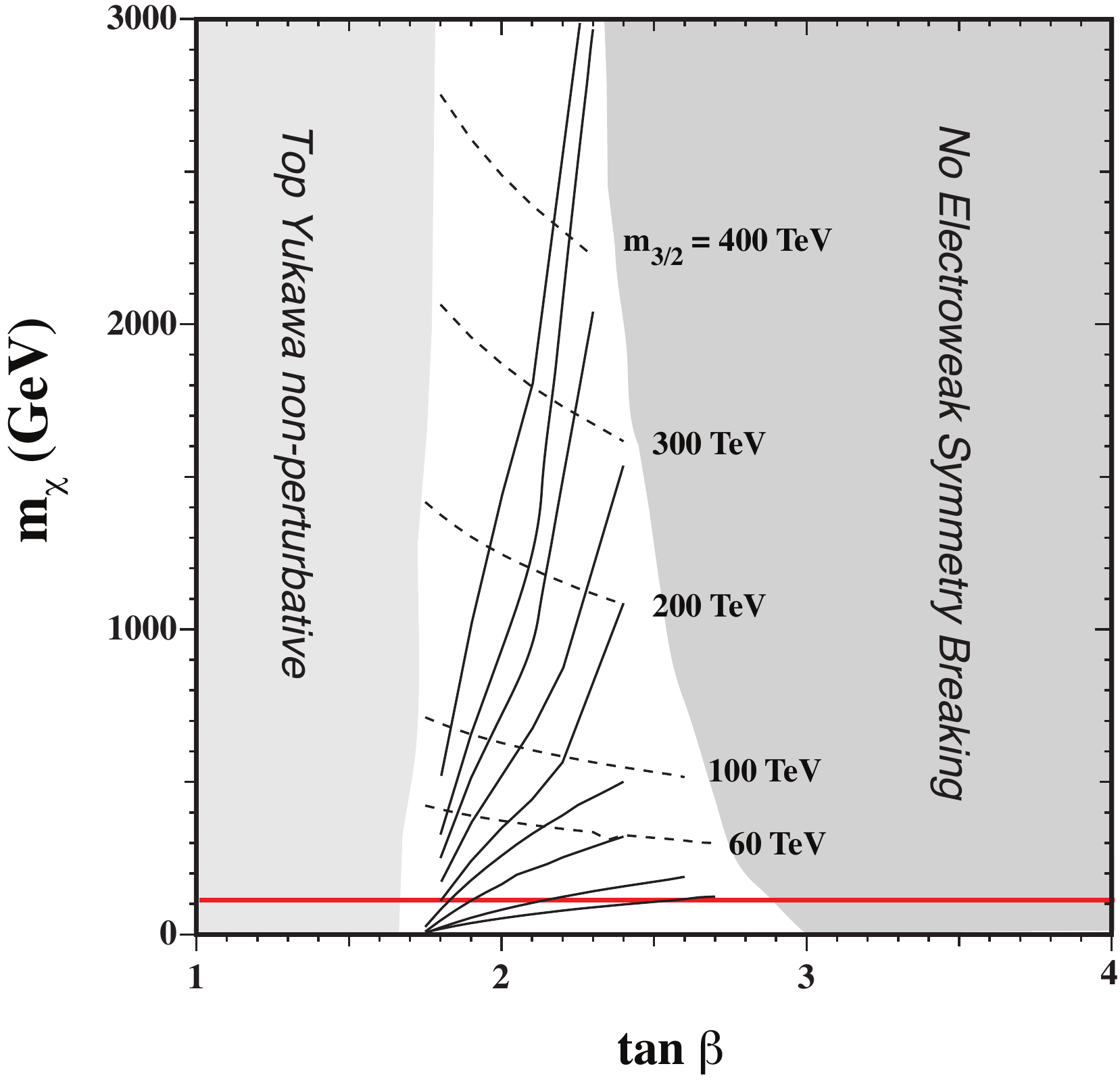}
  \includegraphics[height=.5\textwidth]{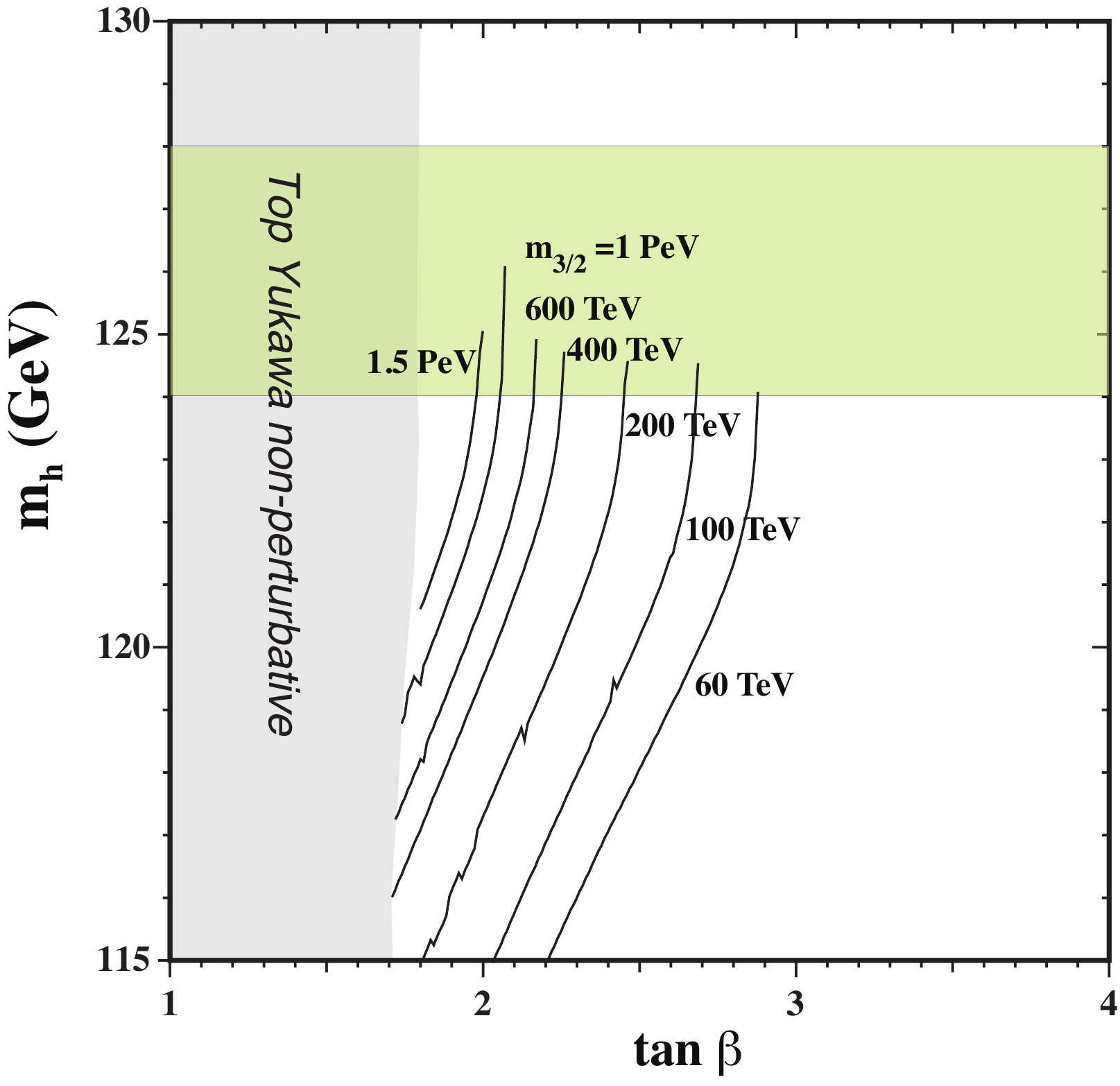}
\caption{
{(left)
The wino mass as a function of $\tan \beta$ for both
$\mu > 0$ (solid) and $\mu < 0$ (dashed).
The LEP bound of $104$\,GeV on the chargino mass is shown as the horizontal red line.
The different curves correspond to different values of $m_{3/2}$. Only the curves with $\mu < 0$
are labelled for clarity. (right) The light Higgs mass as a function of $\tan \beta$.
The LHC range (including an estimate of theoretical uncertainties) of $m_h = 126 \pm 2$ GeV
is shown as the pale green horizontal band.
The different curves correspond to different values of $m_{3/2}$ as marked.
}}
\label{fig:wino}
\end{figure}

In PGM models, the wino is the LSP.  Because of the large threshold corrections the value of the wino mass is strongly dependent on the sign of $\mu$.  For positive values of $\mu$ and small values of $\tan\beta$, the threshold corrections nearly cancel the anomaly mediated contribution. As $\tan\beta$ increases, the threshold corrections shrink and the wino mass approaches the anomaly mediated value. The wino mass is plotted in the left panel of Fig.\,\ref{fig:wino}
as a function of $\tan \beta$  for fixed values of $m_{3/2}$ as labeled.
Solid curves correspond to $\mu > 0$ and dashed curves to $\mu < 0$.
For clarity, only curves for $\mu < 0$ have been labeled.
Large values of $\tan \beta$ are not compatible with radiative electroweak symmetry breaking, 
while perturbativity of the top quark Yukawa coupling excludes the smallest values of $\tan \beta$.
These regions are qualitatively shaded in the left panel of Fig.\,\ref{fig:wino}.

In the right panel of Fig. \ref{fig:wino}, the Higgs mass is shown as a function of $\tan \beta$ for similar
choices of $m_{3/2}$.  We see that for each value of $m_{3/2}$,
the Higgs mass rises as $\tan \beta$ is increased. At some point, the increase is very
sudden as the derived value of $\mu^2$ goes to 0, and we lose the ability to achieve
successful radiative EWSB. As $\mu$ is decreased the Higgsinos are lighter and there is
additional running in the couplings contributing to the Higgs quartic coupling and as a result,
the Higgs mass is highest at this point which corresponds to the focus point
region of the CMSSM \cite{fp}. As one can see, while there is a relatively wide range
of gravitino masses (60-1500) TeV where the Higgs mass is sufficiently large, 
the range of $\tan \beta$ is in contrast highly constrained.  In these PGM models,  the Higgs mass 
is calculated using the procedure outlined in \cite{Giudice:2011cg}.

\begin{figure}[ht!]
 \includegraphics[height=.5\textwidth]{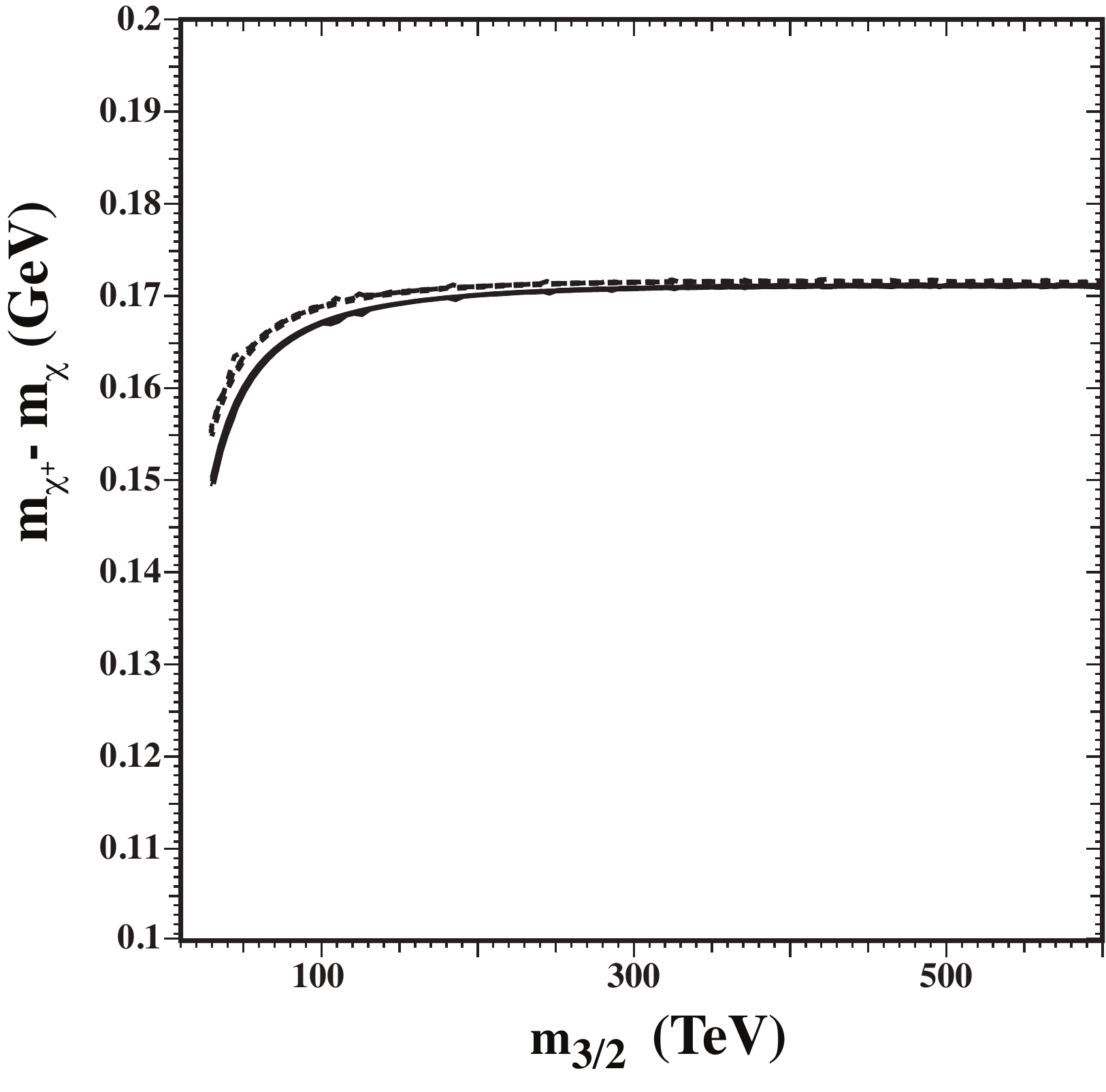}
  \includegraphics[height=.5\textwidth]{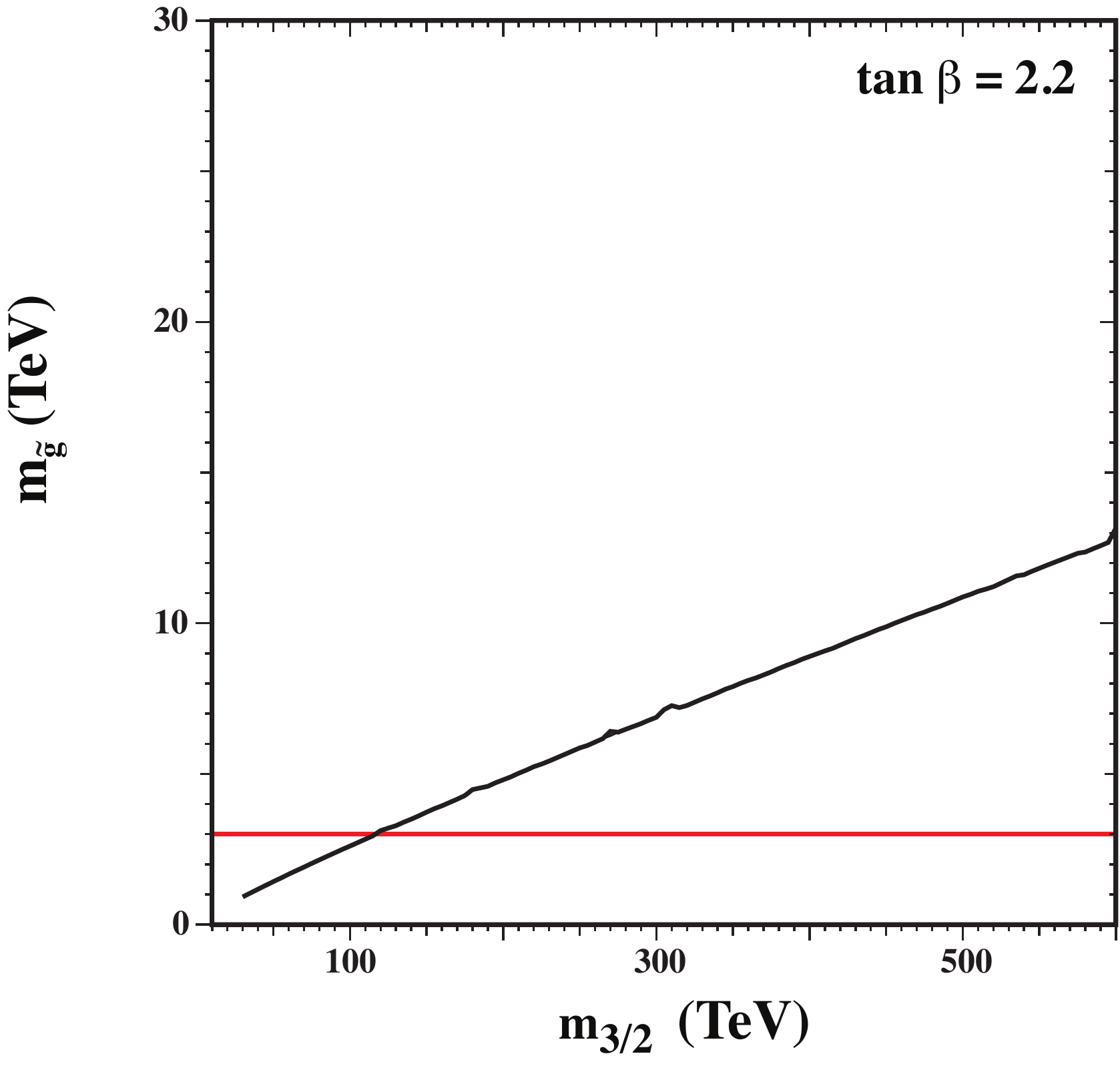}
\caption{
{(left)
The mass difference of the neutral and charged wino as a function of $m_{3/2}$.  The
solid (dashed) curve corresponds to positive (negative) $\mu$. (right) The gluino mass
as a function of $m_{3/2}$. The gluino mass is not sensitive to the sign of $\mu$ and the two curves lie on top of each other. The horizontal line approximates the expected reach of the LHC.
}}
\label{fig:delwino}
\end{figure}

In the left panel of Fig.\,\ref{fig:delwino}, the mass difference between the charged and neutral wino is shown.
The mass difference is just barely larger than the pion mass.
Thus, the decay products of the charged wino will be a very soft pion and the wino LSP.
Because of this very small mass difference, the decay width will have a suppression factor of order $m_{\pi}/m_{\chi}$ and so the chargino will be quite long lived.
The experimental signatures of this decay have been discussed in PGM\,\cite{pgm,pgm2} and related scenarios \cite{ggw,dlmmo}.

In the right panel of  Fig.\,\ref{fig:delwino}, the gluino mass, $m_{\tilde g}$ is shown 
as a function of $m_{3/2}$.
The gluino mass is essentially independent of $\tan \beta$ and here $\tan \beta = 2.2$.
As one might expect, the gluino mass is effectively given by Eq.\,(\ref{eq:M3})
(any deviation is due to two-loop effects which are included in the running).
As summarized in Ref.\,\cite{pgm2}, the most severe limit on the gluino mass
is obtained from the search for multi-jets plus missing transverse energy events
at the LHC, which leads to $m_{\tilde g} > 1.2\,(1.0)$\,TeV
for a wino mass of $100\,(500)\,$GeV \cite{atlas}.%
The red line represents the approximate  reach of the LHC.
A gluino mass of roughly $3$\,TeV corresponds to a gravitino mass of $m_{3/2}=110$\,TeV,
and thus only the low mass end of PGM models may be probed using the gluino.

\begin{figure}[ht!]
\center{ \includegraphics[height=.5\textwidth]{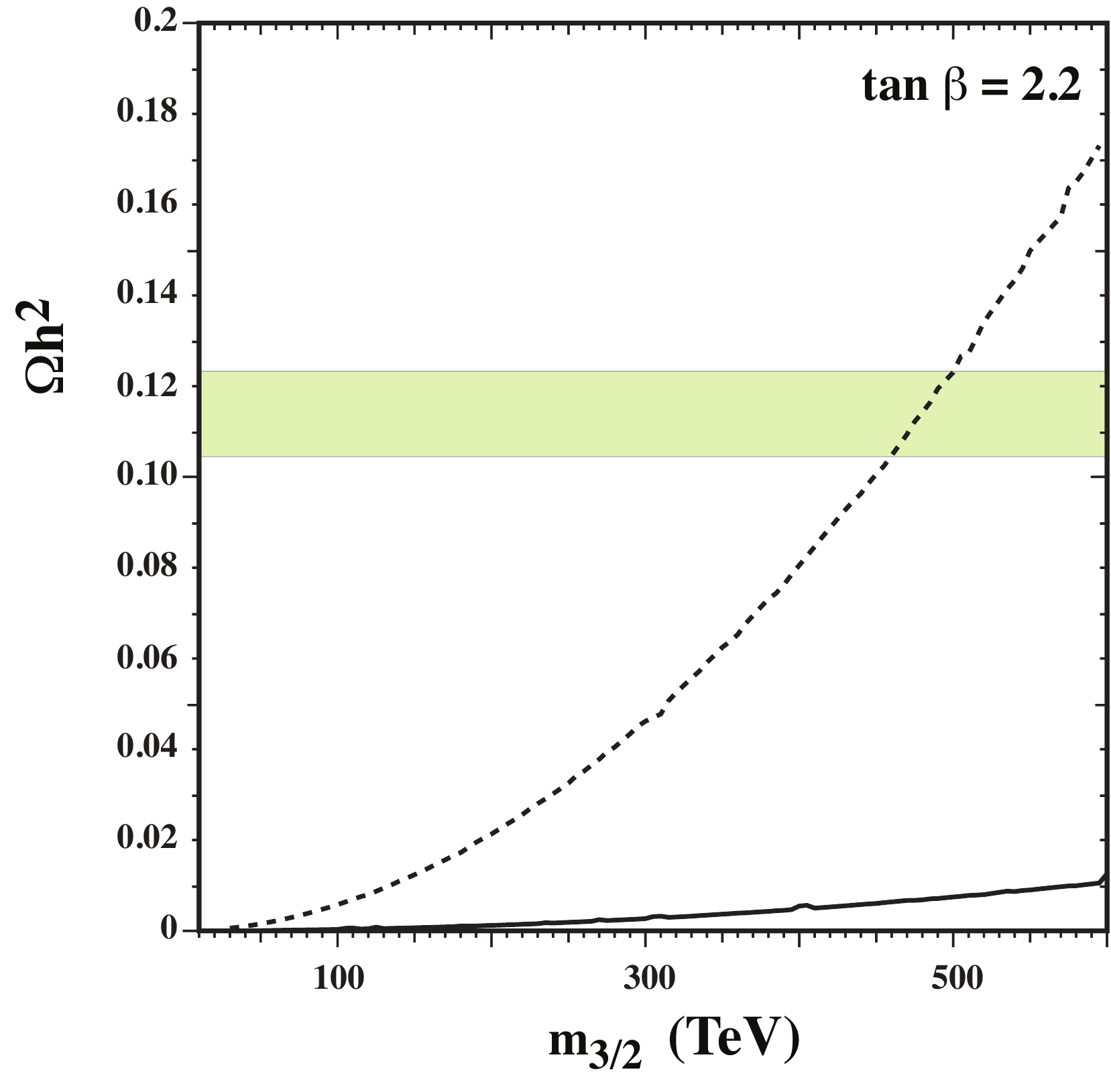}}
\caption{
{$\Omega_\chi h^2$
as a function of $m_{3/2}$ for both
$\mu > 0$ (solid) and $\mu < 0$ (dashed). The Planck range for the relic density is shown by the pale green horizontal band.
 The relic density crosses this band for $\mu <0$ when $m_{3/2} \simeq 460$--$500$\,TeV.
}}
\label{fig:ohsq}
\end{figure}

Finally, we show the dependence of the wino relic density, $\Omega_\chi h^2$ in 
Fig.\,\ref{fig:ohsq} as a function of $m_{3/2}$.
The relic density is also essentially independent of $\tan \beta$ which has been fixed here at 
$\tan \beta = 2.2$.
As one can see, the relic density is always extremely small for $\mu > 0$
and therefore dark matter must either come from some other sector of the theory (e.g. axions)
or there must be a non-thermal source for wino production after freeze-out.
For $\mu < 0$, it is possible to get an acceptable relic density when $m_{3/2}\simeq 460 - 500$\,TeV.
Amazingly, this is consistent with the Higgs mass measurements if the ratio of the Higgs vevs is in the rather small range $\tan\beta =2.1-2.3$.
With $m_{3/2}$ this large, detection of the particle spectrum implied by these models at the LHC is very unlikely.
Even if we abandon hope for discovery at the LHC, the large annihilation cross section
of wino dark matter is in tension with gamma-ray observations of the Galactic center
in the Fermi-LAT and the H.E.S.S. telescope \cite{wino},
although there is still some ambiguity in the dark matter profile at the Galactic center.

\begin{figure}[ht!]
  \includegraphics[height=.5\textwidth]{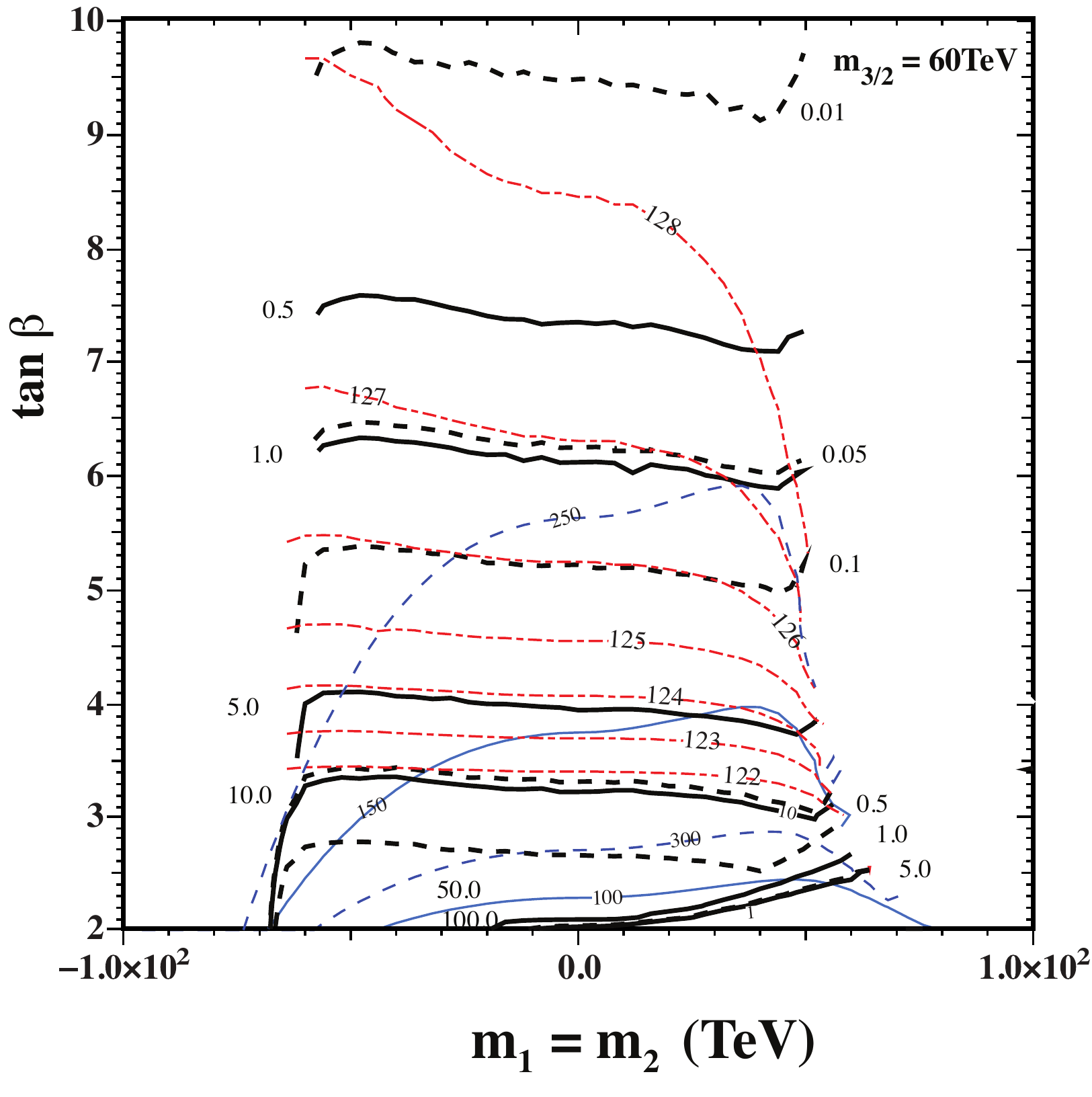}
  \includegraphics[height=.5\textwidth]{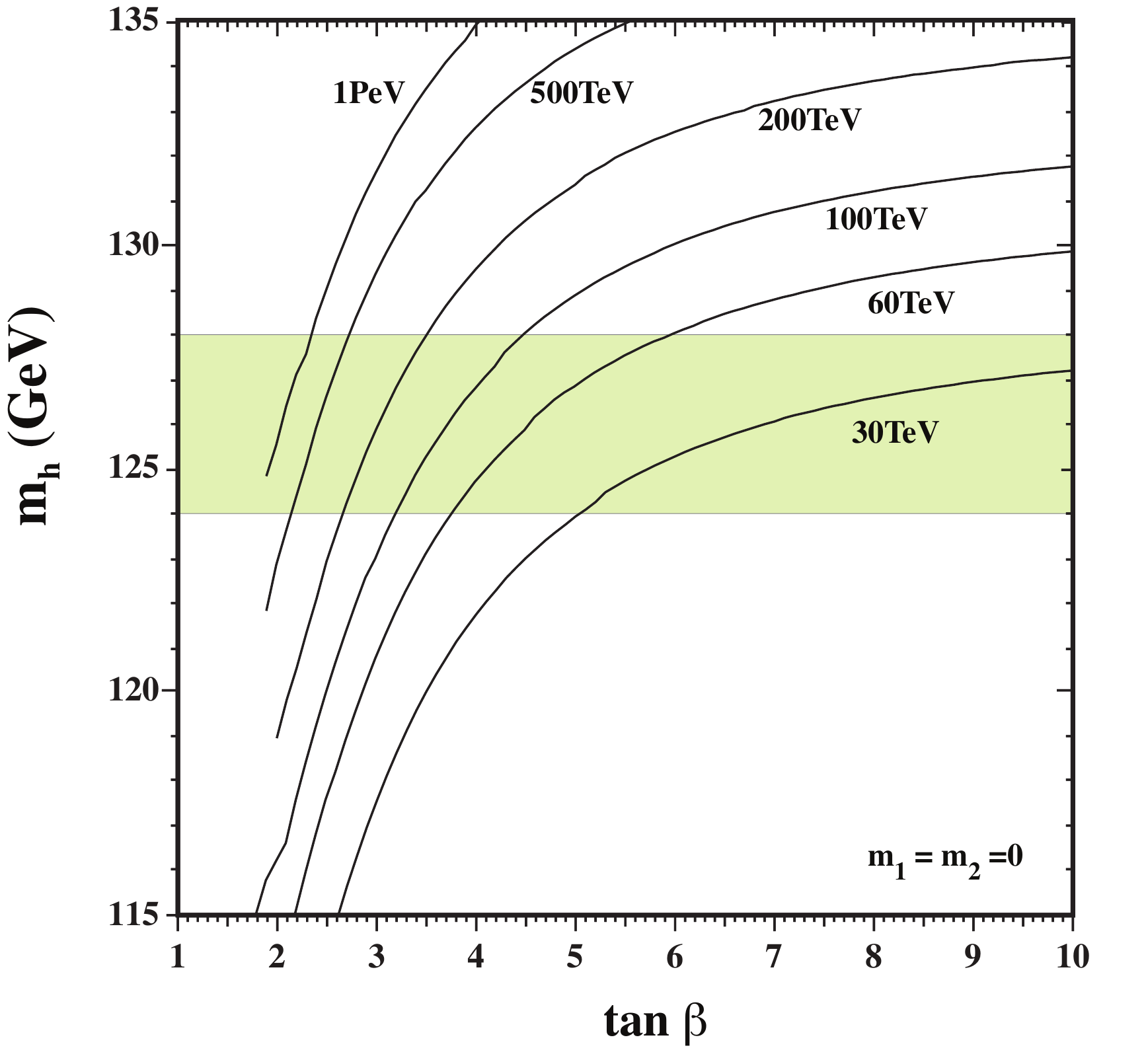}
\caption{
{(left)
The $\tan \beta$--$m_{1,2}$ plane for $m_{3/2} = 60$\,TeV.
The Higgs mass is shown by the nearly horizontal thin red contours in $1$\,GeV intervals.
The wino/chargino mass is shown by the thin solid ($\mu > 0$) and dashed
($\mu < 0$) contours. The thick black contours show the value of the proton lifetime
based on the quark Yukawa couplings (solid) and lepton Yukawa couplings (dashed) in
units of $10^{35}$ years. Lifetime contours for the solid curves are labeled to the left of the contours
whereas dashed contours are labelled to the right. (right) The light Higgs mass as a function of $\tan \beta$ assuming $m_1 = m_2 = 0$
and specific choices of $m_{3/2}$
from $30$--$1000$\,TeV as labeled.
}}
\label{fig:pgmnuhm}
\end{figure}

Somewhat more freedom in the PGM is possible if we allow the Higgs soft masses to be non-Universal \cite{eioy2}. If $m_1 = m_2 \ne m_{3/2}$, we have a 3-parameter PGM theory. 
In the left panel of Fig. \ref{fig:pgmnuhm} contours of the neutralino and Higgs masses
in the $(m_1 = m_2, \tan \beta)$ plane with $m_{3/2} = 60$ TeV are shown. The universal model
discussed above corresponds to $m_1 = m_2 = 60$ TeV and is only valid at low $\tan \beta$. Curves
end when solutions to the electroweak symmetry breaking conditions can not be found.
 Higgs mass contours are shown by red dot-dashed  curves labelled from 122 -128 GeV. 
 Neutralino masses are given by the thin blue curves.  For $\mu> 0$  the curves are solid and the mass increases with $\tan \beta$ (shown are the 100 and 180 GeV contours).  For $\mu<0$  the curves are dashed and the mass decreases with $\tan \beta$ (shown are the 300 and 250 GeV contours). 
 Also shown here are proton decay lifetimes in units of $10^{35}$ yrs.
 Solid curves (labelled to the left) use quark Yukawa couplings, while dashed curves (labelled to the right)
 use lepton Yukawas. For details see \cite{evno}. 

The Higgs mass for special case of $m_1 = m_2 = 0$ is highlighted in the right panel of Fig. \ref{fig:pgmnuhm} which shows the calculated Higgs mass as a function of $\tan \beta$ for several choices
of $m_{3/2}$. 

\begin{figure}[ht!]
 \includegraphics[height=.5\textwidth]{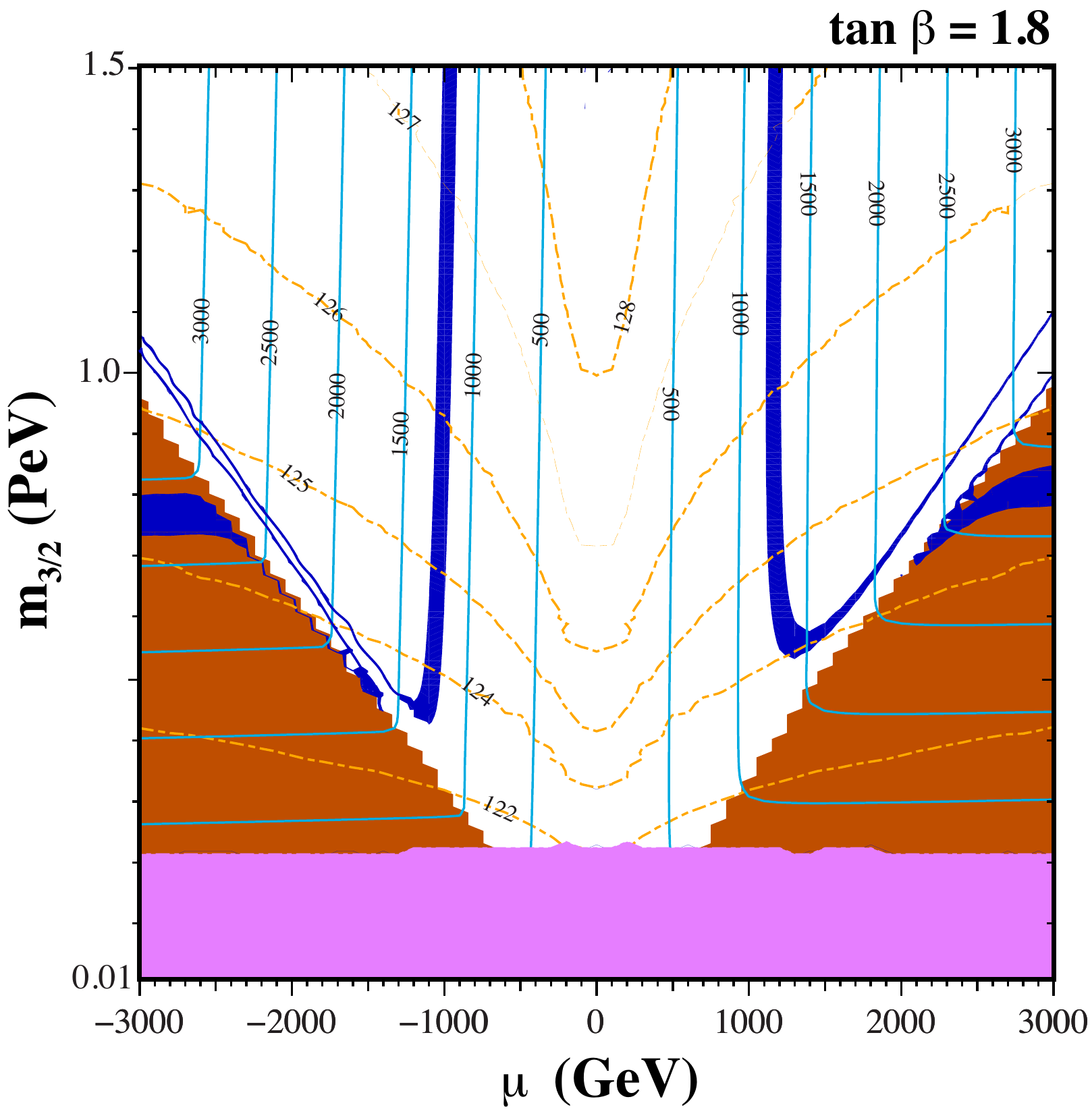}
  \includegraphics[height=.5\textwidth]{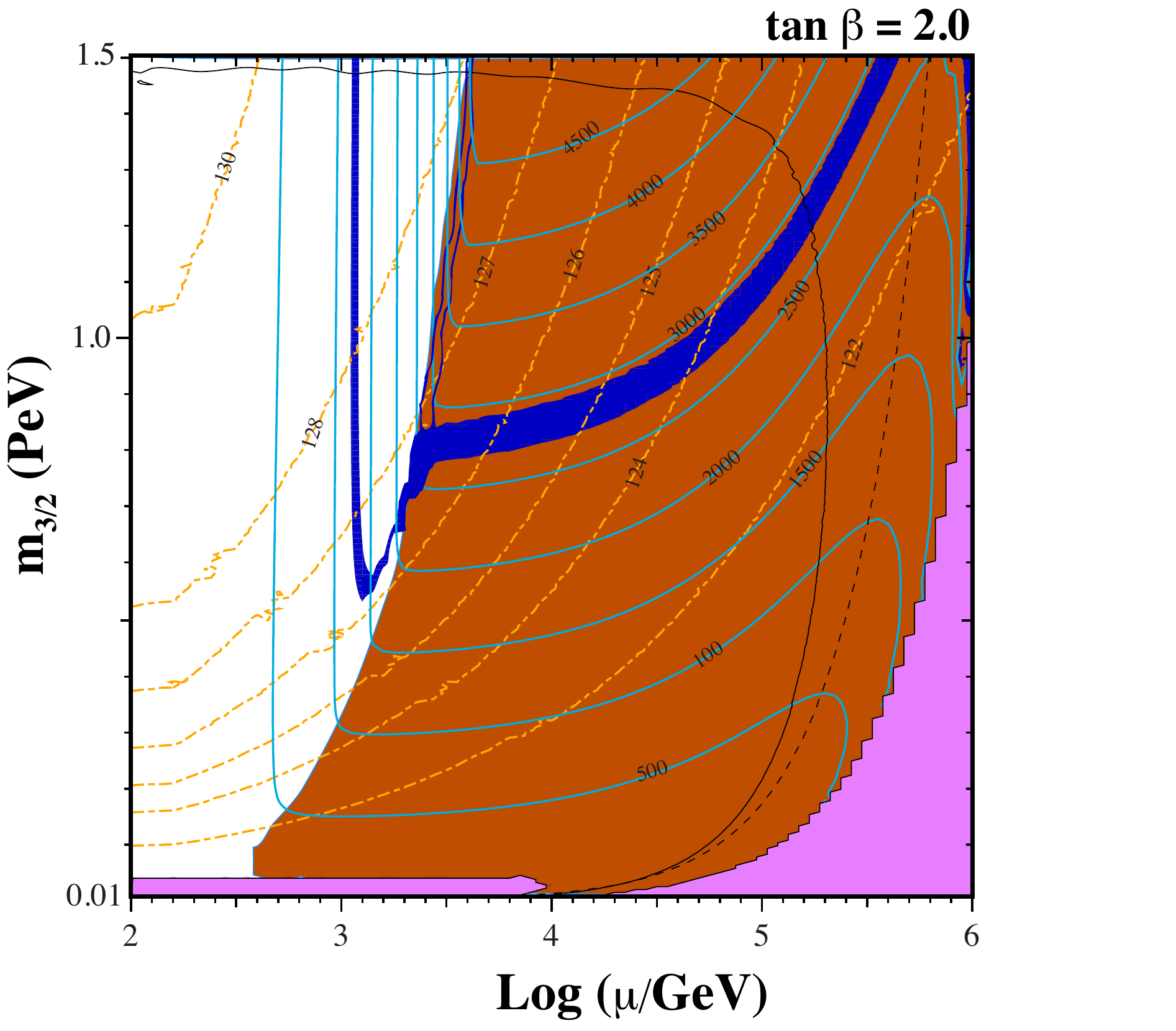}
\caption{
{(left)
The $(\mu, m_{3/2})$ plane for fixed $\tb = 1.8$ (left). The pink shaded region is excluded
as it contains a tachyonic stop. In the dark red shaded region, there is a wino LSP. In the remainder of the plane, the Higgsino is the LSP. Higgs mass contours, shown as orange dot-dashed curves, are given for
$m_H$ = 122, 124, 125, 126, 127, and 128 GeV.  The light blue contours give the
LSP mass from 500-3000 GeV in 500 GeV intervals. When the LSP is a Higgsino, the contours are nearly vertical as the mass of the Higgsino depends primarily on $\mu$, whereas when the LSP is a wino, the contours are nearly horizontal as the anomaly mediated wino mass depends primarily on the gravitino mass.
Strips shaded blue correspond to the correct relic density. In the right panel, the horizontal scale is logarithmic so as to extend to large values of $\mu$.  Here $\tan \beta = 2.0$ was chosen. Included here are also two thin black contours indicating the location
of scalar mass universality ($m_i/m_{3/2} = 1$) (solid) and $c_H = 0$ (dashed).
}}
\label{fig:higgsino}
\end{figure}

As noted earlier, in PGM models with non-universal scalar masses, it is also possible to choose
$\mu$ as an independent parameter rather than the Higgs soft masses. This allows one to more easily
study different aspects of the model with a Higgsino as the LSP (rather than winos) \cite{eioy5}. 
In the left panel of Fig.\,\ref{fig:higgsino}, an example of a $(\mu, m_{3/2})$ plane for fixed $\tb = 1.8$ is shown.
At each point on the plane, the EWSB conditions are used to solve for the $B$ term (or equivalently the GM coupling) and the Higgs soft masses (assumed here to be equal at the GUT scale, as in the NUHM1). The pink shaded region with $m_{3/2} \lesssim 200$ TeV, is excluded as the low energy spectrum contains a tachyonic stop.  In each panel, the dark red shaded regions, delineate the parts of the plane
with a wino LSP. In the complement, there is a Higgsino LSP. There are two sets of contours
in each panel. The orange dot-dashed contours correspond to a constant Higgs mass, and the
light blue contours show the values of the LSP mass across the plane.

In the dark blue shaded regions,
the LSP has a relic density $\Omega h^2 = 0.11 - 0.13$ as preferred by recent Planck results \cite{Planck15}.
The relic density in the Higgsino region depends only on the mass of the Higgsino \cite{osi} and the correct
relic density is obtained when $\mu \approx -1000$ GeV and 1200 GeV, with a Higgsino mass
just over 1200 GeV for both positive and negative $\mu$ (the difference between the Higgsino mass
and $\mu$ is due to one-loop threshold corrections).  These regions are seen as the
vertical strips in both panels.  At  large $|\mu|$, the wino is the LSP and
can have the correct relic density when its mass is approximately 2.7 TeV at $m_{3/2} \approx 800$ TeV.
In between these two extremes, the relic density can be obtained through coannihilation.
These coannihilation strips are seen as diagonal blue strips. For further details on the structure of the 
relic density strips, see \cite{eioy5}. Once again, we see that Higgsino dark matter can be made consistent
with a Higgs mass  of 125 GeV. In this PGM model, this occurs when $m_{3/2} \approx 700$ TeV.

An extended view of the $(\mu, m_{3/2})$ plane for $\mu > 0$ is shown in the right panel of  Fig.\,\ref{fig:higgsino} for $\tb=2.0$.
As in the left panel, the pink shaded region at very low $m_{3/2}$ is excluded due to a tachyonic
stop, and the pink shaded region in the lower right is excluded due to a tachyonic pseudo-scalar.
Once again, we see a strip of Higgsino dark matter with $m_\chi \approx 1200$ GeV running
vertical near $\log \mu = 3.1$ This is connected to a thick strip of wino dark matter which runs towards
high $\mu$ and high $m_{3/2}$ with $m_\chi \lesssim 3$ TeV. In between, we see again two strips near the border
between a Higgsino and wino LSP due to coannihilations.
Also plotted here is the contour where
we have full scalar mass universality.  This is seen as a black curve which is nearly horizontal near
$m_{3/2} = 1.5$ PeV and nearly vertical at $\log \mu \sim 5.2$. The dashed black contour shows the position of the GM coupling $c_H = 0$.

Finally, we consider an extension of PGM models with a single additional $\mathbf{10}$ and $\mathbf{\overline{10}}$ pair.
Because these states are vector-like,
the most general form of the K\"ahler potential will be
\begin{eqnarray}
K=|\mathbf{10}|^2 +|\mathbf{\overline{10}}|^2 + \left(c_{10} (\mathbf{10} \cdot \mathbf{\overline{10}}) +h.c. \right) \, ,
\end{eqnarray}
which includes a GM-like coupling $c_{10}$ that generates a supersymmetric mixing mass
term, $\mu_{10}$, and a supersymmetry-breaking $B$ term for the additional vector-like fields.
The additional fields
also have a gravity-mediated tree-level soft supersymmetry-breaking mass equal to $m_{3/2}$.
Since the $\mathbf{{10}}$ contains fields with the same quantum numbers as SM fields,
the $\mathbf{10}$ ($\mathbf{\overline{10}}$) can be combined into gauge-invariant operators with
$H_2$ ($H_1$). If we impose only gauge symmetries, the most generic contribution to the superpotential is
\begin{eqnarray}
W=y_t' H_2 Q' U' +y_b' H_1\bar Q \bar U  \, ,
\end{eqnarray}
where $Q'$ and $U'$ are from the $\mathbf{10}$ and $\bar Q$ and $\bar U$ are from the $\mathbf{\overline{10}}$.
However, to preserve $R$ symmetry we must take either $y_b'=0$ or $y_t'=0$. Here, we take $y_b'=0$.
The interactions proportional
to $y_t'$ contribute to the beta function of the up Higgs soft mass in a similar way to those controlled by
the top quark Yukawa coupling, $y_t$.
This extended theory now has four parameters:
$m_{3/2}$, $\tan \beta$, $c_{10}$, and $y_t'$.

As in the previously described PGM models, the gaugino masses are generated by anomalies \cite{anom}.
With the addition of the $\mathbf{10}$
and $\mathbf{\overline{10}}$, the anomaly-mediated contributions to the gaugino masses are now
\begin{eqnarray}
    M_{1} &=&
    \frac{48}{5} \frac{g_{1}^{2}}{16 \pi^{2}}
    m_{3/2}\ ,
    \label{eq:M110} \\
    M_{2} &=&
    \frac{g_{2}^{2}}{4 \pi^{2}} m_{3/2}  \ ,
        \label{eq:M210}     \\
    M_{3} &=& 0\ .
    \label{eq:M310}
\end{eqnarray}
In addition, the gauginos then get rather large threshold corrections from the $\mathbf{10}$ and
$\mathbf{\overline{10}}$ when they are integrated out, which is in addition to the large threshold correction
coming from integrating out the Higgsinos: for more details see \cite{eo}.
Since the only contribution to the mass of the gluino comes from the threshold corrections,
it tends to be lighter than in typical PGM models. Hence there are regions where the gluino can
coannihilate with the bino \cite{glu,shafi, hari,eo,deSimone:2014pda,liantao,raza,ELO,eelo},  yielding the possibility of a relatively heavy dark matter candidate \cite{eelo}.

\begin{figure}[ht!]
\includegraphics[height=.5\textwidth]{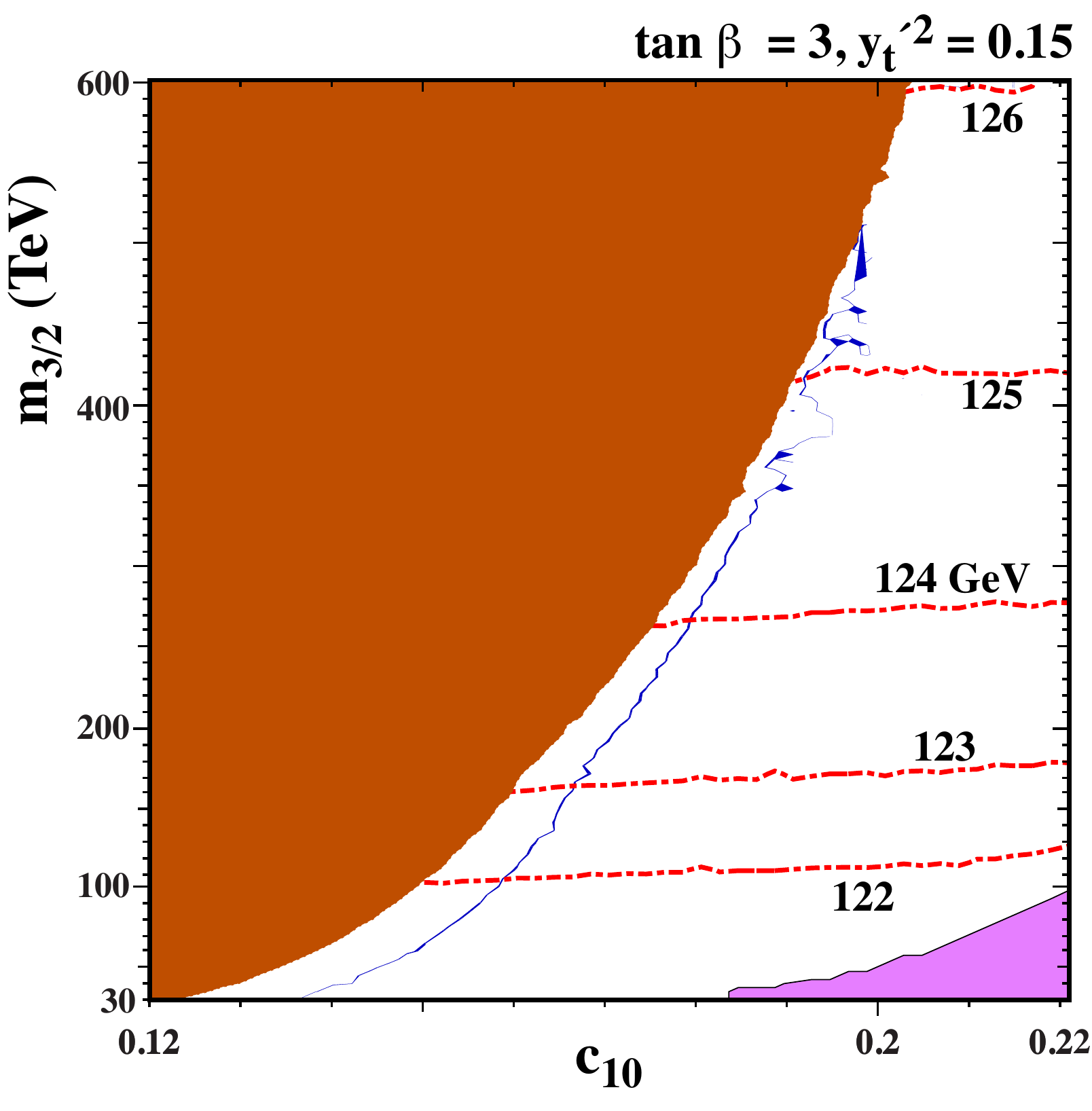} 
\includegraphics[height=.5\textwidth]{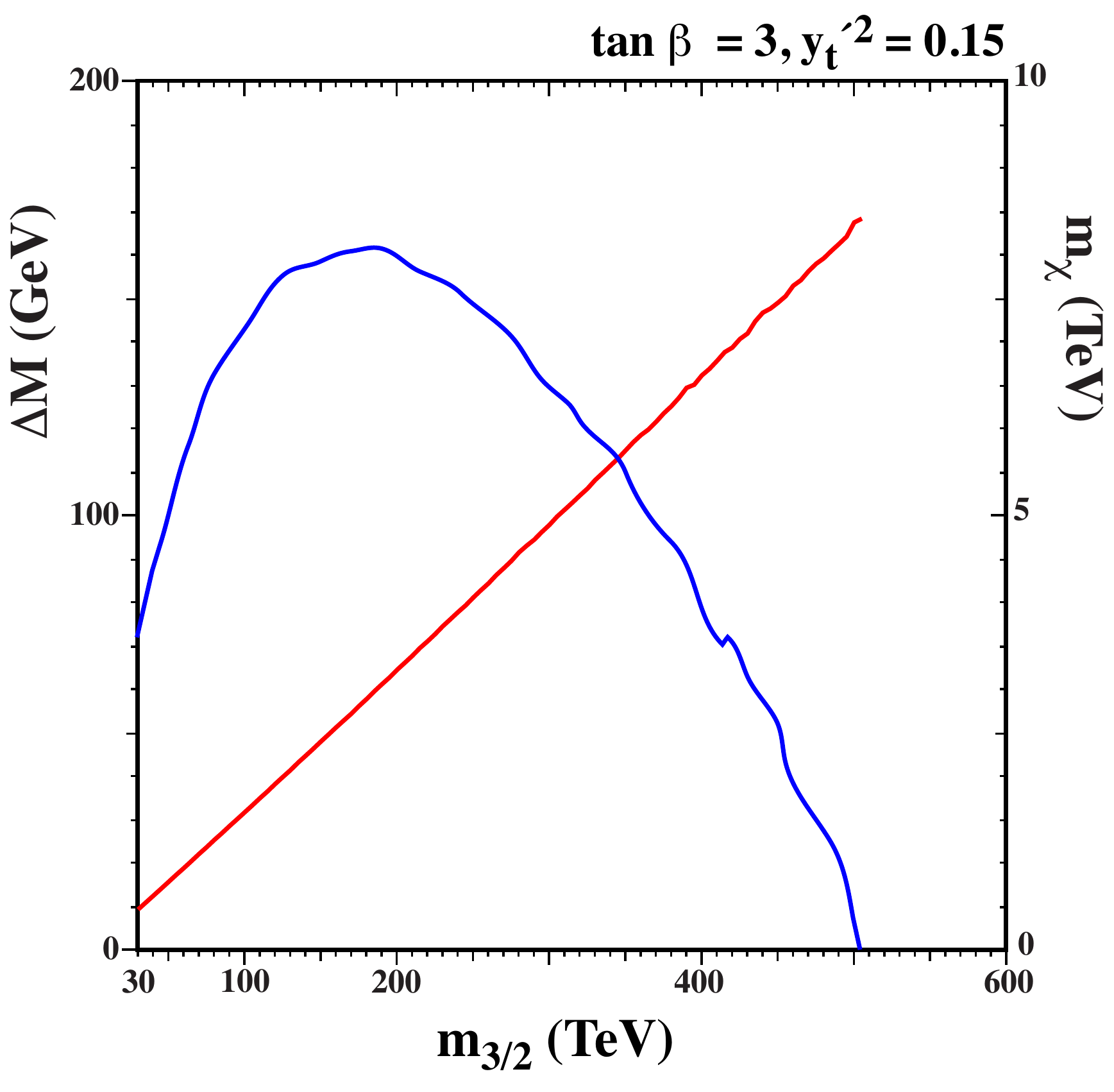}
\caption{\label{fig:EVLSM1}\it
The PGM $(c_H,m_{3/2})$ plane for fixed $\tan \beta = 3$ and $y_t'^2 = 0.15$ \cite{eelo}.
The dark blue strip in the left panel shows where the relic
LSP density $\Omega_\chi h^2$ falls within the $\pm 3$-$\sigma$ range allowed by Planck, 
and the lightest neutralino is no longer the LSP in the low-$c_H$ regions shaded brick-red.
One or more of the new vector scalars becomes tachyonic in the lower right corner of the plane (shaded pink).
The right panel shows the gluino-neutralino mass difference (left axis, blue line) and the neutralino mass (right
axis, red line) as functions of $m_{3/2}$.}
\end{figure}

In Fig. \ref{fig:EVLSM1}, we show the $(c_{10}, m_{3/2})$ plane with
fixed $\tan \beta = 3$ ,and $y_t'^2 = 0.15$. In the left panel, we see
a large red shaded region at low $c_{10}$ where the gluino is the LSP. To the right of this boundary,
we see the gluino coannihilation strip.
In the lower right corner, the pink shaded region is excluded
because one or more of the new vector scalars becomes tachyonic. As in previous figures,
the Higgs mass contours are shown as red dot-dashed curves as labelled. 
In the right panel we see, the gluino-neutralino mass difference $\Delta M$ along the gluino
coannihilation strip (blue) and the neutralino mass along the strip (red).
We see that the curve for $\Delta M$ has the same characteristic shape as we saw above for the stop coannihilation strip, in this case due to strong coannihilations involving the gluino and peaks at $\simeq 170$~GeV  at $m_{3/2} \simeq 200$ TeV when $m_\chi \simeq 3$ TeV.
The end-point of the coannihilation strip occurs at $m_{3/2} \simeq 500$ TeV where $m_\chi \simeq 8.3$ TeV.

\section{Summary}
The discovery of the Higgs boson with $m_h \approx 125$ GeV and the lack of discovery of supersymmetry
at the LHC has put significant pressure on CMSSM-like models. The existence of dark matter
requires physics beyond the standard model, and  supersymmetry offers the possibility
of a dark matter candidate while at the same time ensures the stability of the Higgs potential, allows
for gauge coupling unification, and a solution to the hierarchy problem among its additional benefits.
It was long hoped that low energy supersymmetry could stabilize the hierarchy between
the weak scale and the UV scale given by GUTs or the Planck scale.
Combined with its ability to resolve the discrepancy between theory and experiment regarding $(g_\mu -2)$,
there were great expectations that supersymmetry should have been discovered during run I at the LHC.
The lack of a signal for supersymmetry was nevertheless consistent with the fact that 
Higgs mass was found to be near the upper end of the range predicted in supersymmetric models.
As a consequence, the CMSSM-like models that held so much promise, are now being pushed into
corners of parameter space often requiring a near degeneracy between sparticle masses
to obtain the correct relic density.

Thus while it is becoming increasingly unlikely that supersymmetry can account for the difference
between the GUT scale and the weak scale of O(100) GeV, it may yet account for explaining 
the stability between the GUT scale  and a few TeV or perhaps only to 10-100 TeV or even a PeV.
The large hierarchy problem is still resolved leaving some small amount of fine-tuning left to get to the
scale of $M_Z$. Supersymmetry may still lie within range of future LHC runs or it 
may lie beyond the reach of the LHC
altogether, and require a future higher-energy $pp$ collider for its detection \cite{interplay}.

Here, I have shown several examples of four parameter, CMSSM-like models which remain phenomenologically 
viable. These include the CMSSM with large $A$-terms which allow for sufficiently heavy Higgs masses
while achieving the correct relic density through stop coannihilation. There are several one-parameter
extensions of the CMSSM which offer more viable solutions.  These include the NUHM1 discussed here
where it is relatively straight-forward to construct models with a $\sim 1$ TeV Higgsino LSP \cite{osi}. 
In addition, 
there are subGUT models \cite{subGUT} in which the universality scale is 
reduced below the GUT scale, and models
with a non-Universal gluino mass for which gluino co-annihilations drive the relic density calculation \cite{glu,ELO,eelo}.  It is also possible to construct 2-parameter models such as PGM though
these typically result in a wino LSP which is either underdense or constrained by current 
gamma ray observations. I have also discussed a three-parameter version of PGM with a Higgsino LSP,
and a 4-parameter version with a bino LSP with gluino coannihilations.  

While it is relatively easy to construct any of the models discussed here,
it becomes increasingly more difficult to confirm them with experiment. 

{\bf Acknowledgements} I would like to thank E. Dudas, J. Ellis, J. Evans,  M. Ibe, A. Linde, F. Luo, 
Y. Mambrini, A. Mustafayev, N. Nagata, P. Sandick, 
T. Yanagida and J. Zheng
as well as the Mastercode collaboration for many fruitful collaborations leading to the work described here. 
This work was supported in part by DOE grant DE-SC0011842 at the University of Minnesota.


\begin{thebibliography}{99}

\bibitem{ATLAS20}
G.~Aad {\it et al.}  [ATLAS Collaboration],
  JHEP {\bf 1409} (2014) 176
  [arXiv:1405.7875 [hep-ex]];
  arXiv:1507.05525 [hep-ex];
{full ATLAS Run~1 results can be found at} \\
{\tt https://twiki.cern.ch/twiki/bin/view/AtlasPublic/SupersymmetryPublicResults}.

\bibitem{CMS20}
S.~Chatrchyan {\it et al.}  [CMS Collaboration],
  JHEP {\bf 1406} (2014) 055
  [arXiv:1402.4770 [hep-ex]];
{full CMS Run~1 results can be found at} \\
 {\tt https://twiki.cern.ch/twiki/bin/view/CMSPublic/PhysicsResultsSUS}.
  
    \bibitem{hier}
L.~Maiani,
in Proceedings, Gif-sur-Yvette Summer School On Particle Physics,
  1979, 1-52;
Gerard 't~Hooft and others (eds.),
{\it Recent Developments in Gauge Theories, Proceedings of the Nato Advanced
  Study Institute, Cargese, France, August 26 - September 8, 1979},
Plenum press, New York, USA, 1980, Nato Advanced Study Institutes
  Series: Series B, Physics, 59.;
Edward Witten,
{\em Phys. Lett.} B105, 267, 1981.

\bibitem{Ellis:1990zq}
John~R. Ellis, S.~Kelley and Dimitri~V. Nanopoulos,
{\em Phys. Lett.} B249, 441, 1990;
John~R. Ellis, S.~Kelley and Dimitri~V. Nanopoulos,
{\em Phys. Lett.} B260, 131, 1991;
Ugo Amaldi, Wim de~Boer, and Hermann Furstenau.
\newblock {\em Phys. Lett.}, B260, 447, 1991;
Paul Langacker and Ming-xing Luo,
{\em Phys. Rev.} D44, 817, 1991;
C.~Giunti, C.~W. Kim and U.~W. Lee,
{\em Mod. Phys. Lett.} A6, 1745, 1991.

\bibitem{Ellis:2000ig} 
John~R. Ellis and Douglas Ross,
{\em Phys. Lett.} B506, 331, 2001,  hep-ph/0012067.

\bibitem{ewsb}
L.~E.~Ibanez and G.~G.~Ross,
  Phys.\ Lett.\ B {\bf 110}, 215 (1982);
  K.~Inoue, A.~Kakuto, H.~Komatsu and S.~Takeshita,
  Prog.\ Theor.\ Phys.\  {\bf 68}, 927 (1982)
  [Erratum-ibid.\  {\bf 70}, 330 (1983)]
  [Prog.\ Theor.\ Phys.\  {\bf 70}, 330 (1983)];
L.~E.~Ibanez,
  Phys.\ Lett.\ B {\bf 118}, 73 (1982);
  J.~R.~Ellis, D.~V.~Nanopoulos and K.~Tamvakis,
  Phys.\ Lett.\ B {\bf 121}, 123 (1983);
J.~R.~Ellis, J.~S.~Hagelin, D.~V.~Nanopoulos and K.~Tamvakis,
  Phys.\ Lett.\ B {\bf 125}, 275 (1983);
   L.~Alvarez-Gaume, J.~Polchinski and M.~B.~Wise,
  Nucl.\ Phys.\ B {\bf 221}, 495 (1983).

\bibitem{ehnos} H.~Goldberg,
                Phys.\ Rev.\ Lett.\ {\bf 50} (1983) 1419;
                J.~Ellis, J.~Hagelin, D.~Nanopoulos, K.~Olive and M.~Srednicki,
                Nucl.\ Phys.\ B {\bf 238} (1984) 453.

\bibitem{Ellis:1990nz}
John~R. Ellis, Giovanni Ridolfi and Fabio Zwirner,
{\em Phys. Lett.} B257, 83, 1991;
John~R. Ellis, Giovanni Ridolfi and Fabio Zwirner,
{\em Phys. Lett.} B262, 477, 1991;
Y.~Okada, Masahiro Yamaguchi and T.~Yanagida,
{\em Phys. Lett.} B262, 54, 1991;
Yasuhiro Okada, Masahiro Yamaguchi and Tsutomu Yanagida,
{\em Prog. Theor. Phys.} 85, 1, 1991;
Howard~E. Haber and Ralf Hempfling,
{\em Phys. Rev. Lett.} 66, 1815, 1991.

\bibitem{lhch}
G.~Aad {\it et al.}  [ATLAS Collaboration],
  Phys.\ Lett.\ B {\bf 716}, 1 (2012)
  [arXiv:1207.7214 [hep-ex]];
   S.~Chatrchyan {\it et al.}  [CMS Collaboration],
  Phys.\ Lett.\ B {\bf 716}, 30 (2012)
  [arXiv:1207.7235 [hep-ex]];
G.~Aad {\it et al.}  [ATLAS and CMS Collaborations],
  Phys.\ Rev.\ Lett.\  {\bf 114} (2015) 191803
  [arXiv:1503.07589 [hep-ex]].
  
    \bibitem{funnel}
M.~Drees and M.~M.~Nojiri,
Phys.\ Rev.\ D {\bf 47} (1993) 376 [arXiv:hep-ph/9207234];
H.~Baer and M.~Brhlik,
Phys.\ Rev.\ D {\bf 53} (1996) 597 [arXiv:hep-ph/9508321];
  Phys.\ Rev.\  D {\bf 57} (1998) 567
  [arXiv:hep-ph/9706509];
  H.~Baer, M.~Brhlik, M.~A.~Diaz, J.~Ferrandis, P.~Mercadante, P.~Quintana and X.~Tata,
    Phys.\ Rev.\  D {\bf 63} (2001) 015007
  [arXiv:hep-ph/0005027];
  J.~R.~Ellis, T.~Falk, G.~Ganis, K.~A.~Olive and M.~Srednicki,
Phys.\ Lett.\ B {\bf 510} (2001) 236
[arXiv:hep-ph/0102098].
  
  
  
 \bibitem{cmssm}
 G.~L.~Kane, C.~F.~Kolda, L.~Roszkowski and J.~D.~Wells,
  Phys.\ Rev.\  D {\bf 49} (1994) 6173
  [arXiv:hep-ph/9312272];
  J.~R.~Ellis, T.~Falk, K.~A.~Olive and M.~Schmitt,
Phys.\ Lett.\ B {\bf 388} (1996) 97
[arXiv:hep-ph/9607292];
Phys.\ Lett.\ B {\bf 413} (1997) 355
[arXiv:hep-ph/9705444];
V.~D.~Barger and C.~Kao,
Phys.\ Rev.\ D {\bf 57} (1998) 3131
[arXiv:hep-ph/9704403];
L.~Roszkowski, R.~Ruiz de Austri and T.~Nihei,
JHEP {\bf 0108} (2001) 024
[arXiv:hep-ph/0106334];
A.~Djouadi, M.~Drees and J.~L.~Kneur,
JHEP {\bf 0108} (2001) 055
[arXiv:hep-ph/0107316];
U.~Chattopadhyay, A.~Corsetti and P.~Nath,
Phys.\ Rev.\ D {\bf 66} (2002) 035003
[arXiv:hep-ph/0201001];
J.~R.~Ellis, K.~A.~Olive and Y.~Santoso,
New Jour.\ Phys.\  {\bf 4} (2002) 32
[arXiv:hep-ph/0202110];
H.~Baer, C.~Balazs, A.~Belyaev, J.~K.~Mizukoshi, X.~Tata and Y.~Wang,
JHEP {\bf 0207} (2002) 050
[arXiv:hep-ph/0205325];
R.~Arnowitt and B.~Dutta,
arXiv:hep-ph/0211417.

\bibitem{efgo}
 J.~R.~Ellis, T.~Falk, G.~Ganis, K.~A.~Olive and M.~Schmitt,
  Phys.\ Rev.\ D {\bf 58} (1998) 095002
  [arXiv:hep-ph/9801445];
J.~R.~Ellis, T.~Falk, G.~Ganis and K.~A.~Olive,
  Phys.\ Rev.\ D {\bf 62} (2000) 075010
  [arXiv:hep-ph/0004169].

\bibitem{cmssmwmap}
J.~R.~Ellis, K.~A.~Olive, Y.~Santoso and V.~C.~Spanos,
Phys.\ Lett.\ B {\bf 565} (2003) 176
[arXiv:hep-ph/0303043];
H.~Baer and C.~Balazs,
  JCAP {\bf 0305}, 006 (2003)
  [arXiv:hep-ph/0303114];
A.~B.~Lahanas and D.~V.~Nanopoulos,
  Phys.\ Lett.\  B {\bf 568}, 55 (2003)
  [arXiv:hep-ph/0303130];
U.~Chattopadhyay, A.~Corsetti and P.~Nath,
  Phys.\ Rev.\  D {\bf 68}, 035005 (2003)
  [arXiv:hep-ph/0303201];
   C.~Munoz,
  Int.\ J.\ Mod.\ Phys.\  A {\bf 19}, 3093 (2004)
  [arXiv:hep-ph/0309346];
    R.~Arnowitt, B.~Dutta and B.~Hu,
arXiv:hep-ph/0310103;
   J.~Ellis and K.~A.~Olive,
  arXiv:1001.3651 [astro-ph.CO], published in {\it Particle dark matter}, ed. G.~Bertone, pp. 142-163.

     \bibitem{eo6}
  J.~Ellis and K.~A.~Olive,
  Eur.\ Phys.\ J.\ C {\bf 72}, 2005 (2012)
  [arXiv:1202.3262 [hep-ph]].

 \bibitem{ehow+}
O.~Buchmueller {\it et al.},
  Eur.\ Phys.\ J.\ C {\bf 74} (2014) 3,  2809
  [arXiv:1312.5233 [hep-ph]].

  \bibitem{elos}
  J.~Ellis, F.~Luo, K.~A.~Olive and P.~Sandick,
  Eur.\ Phys.\ J.\ C {\bf 73}, no. 4, 2403 (2013)
  [arXiv:1212.4476 [hep-ph]].
  
  \bibitem{eelnos}
  J.~Ellis, J.~L.~Evans, F.~Luo, N.~Nagata, K.~A.~Olive and P.~Sandick,
  arXiv:1509.08838 [hep-ph].
  
  \bibitem{Fetal}
E.~Cremmer, B.~Julia, J.~Scherk, P.~van Nieuwenhuizen, S.~Ferrara and L.~Girardello,
  Phys.\ Lett.\  B {\bf 79}, 231 (1978);
  E.~Cremmer, B.~Julia, J.~Scherk, S.~Ferrara, L.~Girardello and P.~van Nieuwenhuizen,
  Nucl.\ Phys.\  B {\bf 147}, 105 (1979);
For reviews, see:
H.~P.~Nilles, Phys. Rep. {\bf 110} (1984) 1;
A.~Brignole, L.~E.~Ibanez and C.~Munoz,
arXiv:hep-ph/9707209,
published in {\it Perspectives on supersymmetry}, ed.
G.~L.~Kane, pp. 125-148. 

\bibitem{acn}
 A.~H.~Chamseddine, R.~L.~Arnowitt and P.~Nath,
  Phys.\ Rev.\ Lett.\  {\bf 49}, 970 (1982);
R.~L.~Arnowitt, A.~H.~Chamseddine and P.~Nath,
  Phys.\ Rev.\ Lett.\  {\bf 50}, 232 (1983);
R.~Arnowitt, A.~H.~Chamseddine and P.~Nath,
  Int.\ J.\ Mod.\ Phys.\ A {\bf 27}, 1230028 (2012)
  [Int.\ J.\ Mod.\ Phys.\ A {\bf 27}, 1292009 (2012)]
  [arXiv:1206.3175 [physics.hist-ph]].
  
\bibitem{bfs}
  R.~Barbieri, S.~Ferrara and C.~A.~Savoy,
  Phys.\ Lett.\  B {\bf 119}, 343 (1982).

  \bibitem{vcmssm}
  J.~R.~Ellis, K.~A.~Olive, Y.~Santoso and V.~C.~Spanos,
  Phys.\ Lett.\ B {\bf 573} (2003) 162
  [arXiv:hep-ph/0305212],
  and 
  Phys.\ Rev.\ D {\bf 70} (2004) 055005
  [arXiv:hep-ph/0405110].

\bibitem{nuhm1}
H.~Baer, A.~Mustafayev, S.~Profumo, A.~Belyaev and X.~Tata,
  Phys.\ Rev.\  D {\bf 71}, 095008 (2005)
  [arXiv:hep-ph/0412059];
            H.~Baer, A.~Mustafayev, S.~Profumo, A.~Belyaev and X.~Tata,
               {\em JHEP} {\bf 0507} (2005) 065,
               hep-ph/0504001.

   \bibitem{eosknuhm}
  J.~R.~Ellis, K.~A.~Olive and P.~Sandick,
  Phys.\ Rev.\  D {\bf 78}, 075012 (2008)
  [arXiv:0805.2343 [hep-ph]].


                 \bibitem{nuhm2}
J.~Ellis, K.~Olive and Y.~Santoso,
Phys.\ Lett.\  B~{\bf 539}, 107 (2002)
[arXiv:hep-ph/0204192];
J.~R.~Ellis, T.~Falk, K.~A.~Olive and Y.~Santoso,
Nucl.\ Phys.\ B {\bf 652}, 259 (2003)
[arXiv:hep-ph/0210205].


\bibitem{pgm}
  M.~Ibe, T.~Moroi and T.~T.~Yanagida,
  Phys.\ Lett.\ B {\bf 644}, 355 (2007)
  [hep-ph/0610277];
 M.~Ibe and T.~T.~Yanagida,
  Phys.\ Lett.\ B {\bf 709}, 374 (2012)
  [arXiv:1112.2462 [hep-ph]];
 M.~Ibe, S.~Matsumoto and T.~T.~Yanagida,
  Phys.\ Rev.\ D {\bf 85}, 095011 (2012)
  [arXiv:1202.2253 [hep-ph]]

\bibitem{pgm2}
 B.~Bhattacherjee, B.~Feldstein, M.~Ibe, S.~Matsumoto and T.~T.~Yanagida,
  Phys.\ Rev.\ D {\bf 87}, 015028 (2013)
  [arXiv:1207.5453 [hep-ph]].

\bibitem{ArkaniHamed:2012gw}
  N.~Arkani-Hamed, A.~Gupta, D.~E.~Kaplan, N.~Weiner and T.~Zorawski,
  arXiv:1212.6971 [hep-ph].


  \bibitem{eioy}
  J.~L.~Evans, M.~Ibe, K.~A.~Olive and T.~T.~Yanagida,
  Eur.\ Phys.\ J.\ C {\bf 73}, 2468 (2013)
  [arXiv:1302.5346 [hep-ph]].

  \bibitem{eioy2}
  J.~L.~Evans, K.~A.~Olive, M.~Ibe and T.~T.~Yanagida,
  Eur.\ Phys.\ J.\ C {\bf 73}, 2611 (2013)
  [arXiv:1305.7461 [hep-ph]].

  \bibitem{eo}
  J.~L.~Evans and K.~A.~Olive,
  Phys.\ Rev.\ D {\bf 90}, no. 11, 115020 (2014)
  [arXiv:1408.5102 [hep-ph]].
  
  \bibitem{mc3}
  O.~Buchmueller {\it et al.},
  Eur.\ Phys.\ J.\ C {\bf 64}, 391 (2009)
  [arXiv:0907.5568 [hep-ph]].
  
  \bibitem{mc}
For more information and updates, please see {\tt http://cern.ch/mastercode/}.

\bibitem{mc9}
O.~Buchmueller {\it et al.},
  Eur.\ Phys.\ J.\ C {\bf 74}, no. 6, 2922 (2014)
  [arXiv:1312.5250 [hep-ph]].


  \bibitem{ENOS}
  J.~R.~Ellis, D.~V.~Nanopoulos, K.~A.~Olive and Y.~Santoso,
  Phys.\ Lett.\  B {\bf 633}, 583 (2006)
  [arXiv:hep-ph/0509331].

\bibitem{Planck15}
P.~A.~R.~Ade {\it et al.}  [Planck Collaboration],
  arXiv:1502.01589 [astro-ph.CO].
  
    \bibitem{stau}
  J. Ellis, T. Falk, and K.A. Olive, Phys. Lett.  {\bf B444} (1998) 367
[arXiv:hep-ph/9810360];
J. Ellis, T. Falk, K.A. Olive, and M. Srednicki, {\it Astr. Part. Phys.}
{\bf 13} (2000) 181
[Erratum-ibid.\  {\bf 15} (2001) 413]
[arXiv:hep-ph/9905481];
R.~Arnowitt, B.~Dutta and Y.~Santoso,
Nucl.\ Phys.\ B {\bf 606} (2001) 59
[arXiv:hep-ph/0102181];
M.~E.~G\'omez, G.~Lazarides and C.~Pallis,
Phys. Rev. D {\bf D61} (2000) 123512
[arXiv:hep-ph/9907261];
  Phys.\ Lett. {\bf B487} (2000) 313
[arXiv:hep-ph/0004028];
  Nucl. Phys. B {\bf B638} (2002) 165
[arXiv:hep-ph/0203131];
T.~Nihei, L.~Roszkowski and R.~Ruiz de Austri,
  JHEP {\bf 0207} (2002) 024
[arXiv:hep-ph/0206266];
M.~Citron, J.~Ellis, F.~Luo, J.~Marrouche, K.~A.~Olive and K.~J.~de Vries,
  Phys.\ Rev.\ D {\bf 87}, 036012 (2013)
  [arXiv:1212.2886 [hep-ph]].

  \bibitem{LEPsusy}
  Joint LEP~2 Supersymmetry Working Group,
  {\it Combined LEP Chargino Results, up to 208 GeV}, \\
  {\tt http://lepsusy.web.cern.ch/lepsusy/www/inos{\_}
  moriond01/charginos{\_}pub.html}.

  \bibitem{bsgex}
The Heavy Flavor Averaging Group, D.~Asner {\it et al.}, 
arXiv:1010.1589 [hep-ex], with updates available at 
{\tt http://www.slac.stanford.edu/xorg/}
{\tt hfag/osc/end$\underline{~}$2009.}  
  
  \bibitem{newBNL} [The Muon g-2 Collaboration],
                 {\it Phys. Rev. Lett.} {\bf 92} (2004) 161802, 
                 hep-ex/0401008;
                 G.~Bennett et al.\ [The Muon g-2 Collaboration],
                  {\em Phys.\ Rev.} {\bf D 73} (2006) 072003
                  [arXiv:hep-ex/0602035].

  
 \bibitem{CDMS}   Z.~Ahmed {\it et al.}  [CDMS Collaboration],
  Phys.\ Rev.\ Lett.\  {\bf 102}, 011301 (2009)
  [arXiv:0802.3530 [astro-ph]].

\bibitem{Xe10}   J.~Angle {\it et al.}  [XENON Collaboration],
  Phys.\ Rev.\ Lett.\  {\bf 100}, 021303 (2008)
  [arXiv:0706.0039 [astro-ph]].

\bibitem{xe100100}
E.~Aprile {\it et al.} [XENON100 Collaboration],
  Phys.\ Rev.\ Lett.\  {\bf 107}, 131302 (2011)
  [arXiv:1104.2549 [astro-ph.CO]].
  
  \bibitem{xe100}
   E.~Aprile {\it et al.} [XENON100 Collaboration],
  Phys.\ Rev.\ Lett.\  {\bf 109}, 181301 (2012)
  [arXiv:1207.5988 [astro-ph.CO]].
  
  \bibitem{lux}
   D.~S.~Akerib {\it et al.}  [LUX Collaboration],
  Phys.\ Rev.\ Lett.\  {\bf 112}, 091303 (2014)
  [arXiv:1310.8214 [astro-ph.CO]].
  
  \bibitem{nuback}
    J.~Billard, L.~Strigari and E.~Figueroa-Feliciano,
  Phys.\ Rev.\ D {\bf 89}, no. 2, 023524 (2014)
  [arXiv:1307.5458 [hep-ph]];
P.~Cushman {\it et al.},
arXiv:1310.8327 [hep-ex].


\bibitem{ATLAS5}
 G.~Aad {\it et al.} [ATLAS Collaboration],
  Phys.\ Rev.\ D {\bf 87}, no. 1, 012008 (2013)
  [arXiv:1208.0949 [hep-ex]];
S.~Chatrchyan {\it et al.} [CMS Collaboration],
  JHEP {\bf 1210}, 018 (2012)
  [arXiv:1207.1798 [hep-ex]];
S.~Chatrchyan {\it et al.} [CMS Collaboration],
  Phys.\ Rev.\ Lett.\  {\bf 109}, 171803 (2012)
  [arXiv:1207.1898 [hep-ex]].
  
\bibitem{MH-ATLAS} ATLAS Collaboration, 
                   ATLAS-CONF-2013-014, ATLAS-COM-CONF-2013-025;
CMS Collaboration,
                 CMS-PAS-HIG-13-005.

 \bibitem{stop}
  C.~Boehm, A.~Djouadi and M.~Drees,
  Phys.\ Rev.\  D {\bf 62}, 035012 (2000)
  [arXiv:hep-ph/9911496];
  J.~R.~Ellis, K.~A.~Olive and Y.~Santoso,
  Astropart.\ Phys.\  {\bf 18}, 395 (2003)
  [arXiv:hep-ph/0112113];
  J.~Edsjo, M.~Schelke, P.~Ullio and P.~Gondolo,
  JCAP {\bf 0304}, 001 (2003)
  [hep-ph/0301106];
            J.~L.~Diaz-Cruz, J.~R.~Ellis, K.~A.~Olive and Y.~Santoso,
  JHEP {\bf 0705}, 003 (2007)
  [arXiv:hep-ph/0701229];
  I.~Gogoladze, S.~Raza and Q.~Shafi,
  Phys.\ Lett.\ B {\bf 706}, 345 (2012)
  [arXiv:1104.3566 [hep-ph]];
   M.~A.~Ajaib, T.~Li and Q.~Shafi,
  Phys.\ Rev.\ D {\bf 85}, 055021 (2012)
  [arXiv:1111.4467 [hep-ph]];
   J.~Harz, B.~Herrmann, M.~Klasen, K.~Kovarik and Q.~L.~Boulc'h,
  Phys.\ Rev.\ D {\bf 87} (2013) 5,  054031
  [arXiv:1212.5241];
 J.~Harz, B.~Herrmann, M.~Klasen and K.~Kovarik,
  Phys.\ Rev.\ D {\bf 91} (2015) 3,  034028
  [arXiv:1409.2898 [hep-ph]];
A.~Ibarra, A.~Pierce, N.~R.~Shah and S.~Vogl,
  Phys.\ Rev.\ D {\bf 91}, no. 9, 095018 (2015)
  [arXiv:1501.03164 [hep-ph]].
  

\bibitem{eoz}
 J.~Ellis, K.~A.~Olive and J.~Zheng,
  Eur.\ Phys.\ J.\ C {\bf 74}, 2947 (2014)
  [arXiv:1404.5571 [hep-ph]].  
  
  \bibitem{raza}
    S.~Raza, Q.~Shafi and C.~S.~Ün,
  Phys.\ Rev.\ D {\bf 92}, no. 5, 055010 (2015)
  [arXiv:1412.7672 [hep-ph]].


 \bibitem{fh}
   S.~Heinemeyer, W.~Hollik and G.~Weiglein,
  Eur.\ Phys.\ J.\ C {\bf 9} (1999) 343
  [arXiv:hep-ph/9812472];
  S.~Heinemeyer, W.~Hollik and G.~Weiglein,
  Comput.\ Phys.\ Commun.\  {\bf 124} (2000) 76
  [arXiv:hep-ph/9812320];
   M.~Frank {\it et al.}, 
  JHEP {\bf 0702} (2007) 047
  [arXiv:hep-ph/0611326];
  T.~Hahn, S.~Heinemeyer, W.~Hollik, H.~Rzehak and G.~Weiglein,
  Comput.\ Phys.\ Commun.\  {\bf 180} (2009) 1426.
  see {\tt http://www.feynhiggs.de}~.
  
\bibitem{bmm}
S.~Chatrchyan {\it et al.}  [CMS Collaboration],
 Phys.\ Rev.\ Lett.\  {\bf 111} (2013) 101804
 [arXiv:1307.5025 [hep-ex]];
 R.Aaij {\it et al.}  [LHCb Collaboration],
 Phys.\ Rev.\ Lett.\  {\bf 111} (2013) 101805
 [arXiv:1307.5024 [hep-ex]];
 R.Aaij {\it et al.}  [LHCb and CMS Collaborations],
LHCb-CONF-2013-012, CMS PAS BPH-13-007 (2013).

  
   \bibitem{nonu}
    D.~Matalliotakis and H.~P.~Nilles,
  Nucl.\ Phys.\  B {\bf 435} (1995) 115
  [arXiv:hep-ph/9407251];
  M.~Olechowski and S.~Pokorski,
  Phys.\ Lett.\ B {\bf 344}, 201 (1995)
  [arXiv:hep-ph/9407404];
V.~Berezinsky, A.~Bottino, J.~Ellis, N.~Fornengo,
               G.~Mignola and S.~Scopel,
               {\em Astropart.\ Phys.}  {\bf 5} (1996) 1,
               hep-ph/9508249;
               M.~Drees, M.~Nojiri, D.~Roy and Y.~Yamada,
               {\em Phys.\ Rev.} {\bf D 56} (1997) 276,
               [Erratum-ibid.\ {\bf D 64} (1997) 039901],
               hep-ph/9701219;
               M.~Drees, Y.~Kim, M.~Nojiri, D.~Toya, K.~Hasuko and
               T.~Kobayashi,
               {\em Phys.\ Rev.} {\bf D 63} (2001) 035008,
               hep-ph/0007202;
               P.~Nath and R.~Arnowitt,
               {\em Phys.\ Rev.} {\bf D 56} (1997) 2820,
               hep-ph/9701301;
               A.~Bottino, F.~Donato, N.~Fornengo and S.~Scopel,
               {\em Phys.\ Rev.} {\bf D 63} (2001) 125003,
               hep-ph/0010203;
               S.~Profumo,
               {\em Phys.\ Rev.} {\bf D 68} (2003) 015006,
               hep-ph/0304071;
               D.~Cerdeno and C.~Munoz,
               {\em JHEP} {\bf 0410} (2004) 015,
               hep-ph/0405057.

  
  
  \bibitem{subGUT}
J.~R.~Ellis, K.~A.~Olive and P.~Sandick,
  Phys.\ Lett.\ B {\bf 642}, 389 (2006)
  [hep-ph/0607002];
  J.~R.~Ellis, K.~A.~Olive and P.~Sandick,
  JHEP {\bf 0706}, 079 (2007)
  [arXiv:0704.3446 [hep-ph]];
  J.~R.~Ellis, K.~A.~Olive and P.~Sandick,
  JHEP {\bf 0808}, 013 (2008)
  [arXiv:0801.1651 [hep-ph]].
  
     \bibitem{superGUT}
 L.~Calibbi, Y.~Mambrini and S.~K.~Vempati,
  JHEP {\bf 0709}, 081 (2007)
  [arXiv:0704.3518 [hep-ph]];
  L.~Calibbi, A.~Faccia, A.~Masiero and S.~K.~Vempati,
  Phys.\ Rev.\  D {\bf 74}, 116002 (2006)
  [arXiv:hep-ph/0605139];
  E.~Carquin, J.~Ellis, M.~E.~Gomez, S.~Lola and J.~Rodriguez-Quintero,
  JHEP {\bf 0905} (2009) 026
  [arXiv:0812.4243 [hep-ph]];
J.~Ellis, A.~Mustafayev and K.~A.~Olive,
  Eur.\ Phys.\ J.\ C {\bf 69}, 201 (2010)
  [arXiv:1003.3677 [hep-ph]];
  J.~Ellis, A.~Mustafayev and K.~A.~Olive,
  Eur.\ Phys.\ J.\ C {\bf 69}, 219 (2010)
  [arXiv:1004.5399 [hep-ph]];
  J.~Ellis, A.~Mustafayev and K.~A.~Olive,
  Eur.\ Phys.\ J.\ C {\bf 71}, 1689 (2011)
  [arXiv:1103.5140 [hep-ph]].

\bibitem{dlmmo}
 E.~Dudas, A.~Linde, Y.~Mambrini, A.~Mustafayev and K.~A.~Olive,
  Eur.\ Phys.\ J.\ C {\bf 73}, no. 1, 2268 (2013)
  [arXiv:1209.0499 [hep-ph]].
  
     \bibitem{osi}
  K.~A.~Olive and M.~Srednicki,
  Phys.\ Lett.\ B {\bf 230}, 78 (1989);
    K.~A.~Olive and M.~Srednicki,
  Nucl.\ Phys.\ B {\bf 355}, 208 (1991).

\bibitem{Abe:2014mwa}
  K.~Abe {\it et al.}  [Super-Kamiokande Collaboration],
  Phys.\ Rev.\ D {\bf 90}, 072005 (2014)
  [arXiv:1408.1195 [hep-ex]].

  
  \bibitem{anom}
   M.~Dine and D.~MacIntire,
  Phys.\ Rev.\ D {\bf 46}, 2594 (1992)
  [hep-ph/9205227];
   L.~Randall and R.~Sundrum,
  Nucl.\ Phys.\  B {\bf 557}, 79 (1999)
  [arXiv:hep-th/9810155];
   G.~F.~Giudice, M.~A.~Luty, H.~Murayama and R.~Rattazzi,
  JHEP {\bf 9812}, 027 (1998)
  [arXiv:hep-ph/9810442];
    J.~A.~Bagger, T.~Moroi and E.~Poppitz,
  JHEP {\bf 0004}, 009 (2000)
  [arXiv:hep-th/9911029];
  P.~Binetruy, M.~K.~Gaillard and B.~D.~Nelson,
  Nucl.\ Phys.\  B {\bf 604}, 32 (2001)
  [arXiv:hep-ph/0011081].

 \bibitem{split}
  J.~D.~Wells,
  hep-ph/0306127;
      N.~Arkani-Hamed and S.~Dimopoulos,
  JHEP {\bf 0506}, 073 (2005)
  [arXiv:hep-th/0405159];
G.~F.~Giudice and A.~Romanino,
  Nucl.\ Phys.\  B {\bf 699}, 65 (2004)
  [Erratum-ibid.\  B {\bf 706}, 65 (2005)]
  [arXiv:hep-ph/0406088];
 N.~Arkani-Hamed, S.~Dimopoulos, G.~F.~Giudice and A.~Romanino,
  Nucl.\ Phys.\  B {\bf 709}, 3 (2005)
  [arXiv:hep-ph/0409232];
  J.~D.~Wells,
  Phys.\ Rev.\  D {\bf 71}, 015013 (2005)
  [arXiv:hep-ph/0411041].
  
       \bibitem{gm}
   G.~F.~Giudice and A.~Masiero,
  Phys.\ Lett.\  B {\bf 206}, 480 (1988).

   \bibitem{ikyy}
  K.~Inoue, M.~Kawasaki, M.~Yamaguchi and T.~Yanagida,
  Phys.\ Rev.\ D {\bf 45}, 328 (1992).

    \bibitem{dmmo}
E.~Dudas, Y.~Mambrini, A.~Mustafayev and K.~A.~Olive,
  Eur.\ Phys.\ J.\ C {\bf 72}, 2138 (2012)
  [arXiv:1205.5988 [hep-ph]].

    \bibitem{lmo}
   A.~Linde, Y.~Mambrini and K.~A.~Olive,
  Phys.\ Rev.\ D {\bf 85}, 066005 (2012)
  [arXiv:1111.1465 [hep-th]];
    E.~Dudas, C.~Papineau and S.~Pokorski,
  JHEP {\bf 0702}, 028 (2007)
  [hep-th/0610297];
  H.~Abe, T.~Higaki, T.~Kobayashi and Y.~Omura,
  Phys.\ Rev.\ D {\bf 75}, 025019 (2007)
  [hep-th/0611024].

  
  \bibitem{pol}
J. Polonyi, Budapest preprint KFKI-1977-93 (1977).

 \bibitem{dine}
  M.~Dine, R.~Kitano, A.~Morisse and Y.~Shirman,
  Phys.\ Rev.\ D {\bf 73}, 123518 (2006)
  [hep-ph/0604140];
  R.~Kitano,
  Phys.\ Lett.\ B {\bf 641}, 203 (2006)
  [hep-ph/0607090].
  
   \bibitem{fp}
 J.~L.~Feng, K.~T.~Matchev and T.~Moroi,
  Phys.\ Rev.\ Lett.\  {\bf 84}, 2322 (2000)
  [arXiv:hep-ph/9908309];
  Phys.\ Rev.\  D {\bf 61}, 075005 (2000)
  [arXiv:hep-ph/9909334];
  J.~L.~Feng, K.~T.~Matchev and F.~Wilczek,
  Phys.\ Lett.\  B {\bf 482}, 388 (2000)
  [arXiv:hep-ph/0004043];
  H.~Baer, T.~Krupovnickas, S.~Profumo and P.~Ullio,
  JHEP {\bf 0510} (2005) 020
  [hep-ph/0507282].


\bibitem{Giudice:2011cg}
  G.~F.~Giudice and A.~Strumia,
  Nucl.\ Phys.\ B {\bf 858}, 63 (2012)
  [arXiv:1108.6077 [hep-ph]];
  E.~Bagnaschi, G.~F.~Giudice, P.~Slavich and A.~Strumia,
  JHEP {\bf 1409}, 092 (2014)
  [arXiv:1407.4081 [hep-ph]].

\bibitem{ggw}
  T.~Gherghetta, G.~F.~Giudice and J.~D.~Wells,
  Nucl.\ Phys.\  B {\bf 559}, 27 (1999)
  [arXiv:hep-ph/9904378];
  J.~L.~Feng, T.~Moroi, L.~Randall, M.~Strassler and S.~-f.~Su,
  Phys.\ Rev.\ Lett.\  {\bf 83}, 1731 (1999)
  [hep-ph/9904250];
S.~Asai, T.~Moroi and T.~T.~Yanagida,
 Phys.\ Lett.\ B {\bf 664}, 185 (2008)
 [arXiv:0802.3725 [hep-ph]];
H.~Baer, J.~K.~Mizukoshi and X.~Tata, 
 Phys.\ Lett.\ B {\bf 488}, 367 (2000)
 [hep-ph/0007073];
A.~J.~Barr, C.~G.~Lester, M.~A.~Parker, B.~C.~Allanach and P.~Richardson, 
JHEP {\bf 0303}, 045 (2003)  [hep-ph/0208214]
 
 \bibitem{atlas}
  The ATLAS Collaboration, ATLAS-CONF-2012-109.

   \bibitem{wino}
  T.~Cohen, M.~Lisanti, A.~Pierce and T.~R.~Slatyer,
  JCAP {\bf 1310}, 061 (2013)
  [arXiv:1307.4082];
J.~Fan and M.~Reece,
  JHEP {\bf 1310}, 124 (2013)
  [arXiv:1307.4400 [hep-ph]];
 M.~Baumgart, I.~Z.~Rothstein and V.~Vaidya,
  JHEP {\bf 1504}, 106 (2015)
  [arXiv:1412.8698 [hep-ph]].
  
  \bibitem{evno}
  J.~L.~Evans, N.~Nagata and K.~A.~Olive,
  Phys.\ Rev.\ D {\bf 91}, 055027 (2015)
  [arXiv:1502.00034 [hep-ph]].
  
  \bibitem{eioy5}
  J.~L.~Evans, M.~Ibe, K.~A.~Olive and T.~T.~Yanagida,
  Phys.\ Rev.\ D {\bf 91}, 055008 (2015)
  [arXiv:1412.3403 [hep-ph]].
  
  
    \bibitem{glu}
 S.~Profumo and C.~E.~Yaguna,
  Phys.\ Rev.\ D {\bf 69}, 115009 (2004)
  [hep-ph/0402208];
 D.~Feldman, Z.~Liu and P.~Nath,
  Phys.\ Rev.\ D {\bf 80}, 015007 (2009)
  [arXiv:0905.1148 [hep-ph]];
N.~Chen, D.~Feldman, Z.~Liu, P.~Nath and G.~Peim,
  Phys.\ Rev.\ D {\bf 83}, 035005 (2011)
  [arXiv:1011.1246 [hep-ph]].

\bibitem{shafi}
I.~Gogoladze, R.~Khalid and Q.~Shafi,
  Phys.\ Rev.\ D {\bf 79}, 115004 (2009)
  [arXiv:0903.5204 [hep-ph]];
I.~Gogoladze, R.~Khalid and Q.~Shafi,
  Phys.\ Rev.\ D {\bf 80}, 095016 (2009)
  [arXiv:0908.0731 [hep-ph]];
  M.~Adeel Ajaib, T.~Li, Q.~Shafi and K.~Wang,
  JHEP {\bf 1101}, 028 (2011)
  [arXiv:1011.5518 [hep-ph]].

 \bibitem{hari}
  K.~Harigaya, M.~Ibe and T.~T.~Yanagida,
  JHEP {\bf 1312}, 016 (2013)
  [arXiv:1310.0643 [hep-ph]];
 K.~Harigaya, K.~Kaneta and S.~Matsumoto,
  Phys.\ Rev.\ D {\bf 89}, no. 11, 115021 (2014)
  [arXiv:1403.0715 [hep-ph]].

      \bibitem{deSimone:2014pda}
   A.~De Simone, G.~F.~Giudice and A.~Strumia,
  JHEP {\bf 1406}, 081 (2014)
  [arXiv:1402.6287 [hep-ph]].
  
    \bibitem{liantao}
    M.~Low and L.~T.~Wang,
  JHEP {\bf 1408}, 161 (2014)
  [arXiv:1404.0682 [hep-ph]];
 N.~Nagata, H.~Otono and S.~Shirai,
  Phys.\ Lett.\ B {\bf 748} (2015) 24
  [arXiv:1504.00504 [hep-ph]].

  
   \bibitem{ELO}
 J.~Ellis, F.~Luo and K.~A.~Olive,
  JHEP {\bf 1509}, 127 (2015)
  [arXiv:1503.07142 [hep-ph]].


\bibitem{eelo}
J.~Ellis, J.~L.~Evans, F.~Luo and K.~A.~Olive,
  arXiv:1510.03498 [hep-ph].

  \bibitem{interplay}
O.~Buchmueller, M.~Citron, J.~Ellis, S.~Guha, J.~Marrouche, K.~A.~Olive, K.~de Vries and J.~Zheng,
  Eur.\ Phys.\ J.\ C {\bf 75} (2015) 10,  469
  [arXiv:1505.04702 [hep-ph]].


 \end{thebibliography}
\end{document}